\documentclass[11pt]{article}
\usepackage{mydef2col}
\usepackage{array}
\usepackage{kantlipsum,widetext}
\usepackage[T1]{fontenc}
\usepackage{bm}
\usepackage[autostyle=true]{csquotes}

\let\svthefootnote\thefootnote
\providecommand{\keywords}[1]{\let\thefootnote\relax\footnotetext{\small \textbf{Keywords:} #1}}
\providecommand{\codes}[1]{\let\thefootnote\relax\footnotetext{\small \textbf{2010 MSC:} #1}}

\allowdisplaybreaks[4]
\graphicspath{{./Graphs/}}
\usepackage{tikz}
\usetikzlibrary{calc,decorations.markings}
\usetikzlibrary{arrows,patterns,positioning}

\usepackage[round]{natbib}

\def\baru{{\bar{u}}}
\def\barw{{\bar{w}}}

\def\Jnu{{J_{|\nu|}}}

\def\hx{{\hat x}}

\newcommand{\iu}{\mathrm{i}\mkern1mu}

\title{Semi-analytical pricing of barrier options in the time-dependent $\lambda$-SABR model.}

\author{
\authorstyle{
Andrey Itkin{}
\textsuperscript{1}
and Dmitry Muravey
\textsuperscript{2}
}
\newline\newline
\textsuperscript{1}
\institution{Tandon School of Engineering, New York University, 1 Metro Tech Center, 10th floor, Brooklyn NY 11201, USA} \\
\textsuperscript{2}
\institution{Moscow State University, Moscow, Russia}
}

\vspace{-0.2in}
\date{}
\begin{document}

\maketitle
\vspace{-0.2in}

\lettrineabstract{We extend the approach of Carr, Itkin and Muravey, 2021 for getting semi-analytical prices of barrier options for the time-dependent Heston model with time-dependent barriers by applying it to the so-called $\lambda$-SABR stochastic volatility model. In doing so we modify the general integral transform method (see Itkin, Lipton, Muravey, Generalized integral transforms  in mathematical finance, World Scientific, 2021) and deliver solution of this problem in the form of Fourier-Bessel series. The weights  of this series solve a linear mixed Volterra-Fredholm equation (LMVF) of the second kind also derived in the paper.  Numerical examples illustrate speed and accuracy of our method which are comparable with those of the finite-difference approach at small maturities and outperform them at high maturities even by using a simplistic implementation of the RBF method for solving the LMVF.}
\hspace{10pt}


\codes{91G40; 91G60; 91G80; 47G30; 35R09; 65L12;}

\let\thefootnote\svthefootnote


\section*{Introduction} \label{layer}

The SABR model is one of the most popular models in mathematical finance. Introduced in \citep{hagan2002} (and then an arbitrage-free version in \citep{SABR2014}), it quickly became a standard tool among practitioners especially in the interest rate derivative markets due to its ability to
capture skew and smile features observed in the interest rates implied volatilities. For European options it has an asymptotic solution, \citep{hagan2002} which allows closed form representation of the implied volatility. Later a SABR  extension for negative interest rates (the shifted SABR model) was developed as well as advanced analytics was provided by using a heat kernel expansion technique, see \citep{Antonov2019} and references therein. As mentioned in that book, since the SABR model allows an analytic approximation for Black volatilities it is widely used for the swaption volatility cube interpolation and extrapolation.

The SABR model can also be extended to its time-dependent version where all model parameters are deterministic functions of time. This obviously complicates the calibration procedure. An advanced calibration method of the time-dependent SABR model based on so-called "effective parameters" was developed in \citep{Oosterlee2015}.

Another extension was proposed in \citep{Hagan2020}.  The authors consider the mean-reverting SABR model where the stochastic process for the instantaneous volatility has a mean-reverting drift. They also assume that coefficients of the model are functions of time. Then they develop asymptotic methods to obtain an effective forward equation for the marginal density of the forward price. This equation is not exact, but as accurate as the SABR implied volatility formulas. The authors suggest solving this one-dimensional partial differential equation numerically to obtain the density. European option prices can then be found by integrating this density with the option payoff.

As far as exotic options is concerned, for the standard SABR model with constant coefficients in \citep{YangLiuCui2017}
a closed form expression is obtained to approximate prices of various types of barrier option (Down-and-Out/In, Up-and-Out/In). The authors derive an approximate formula for the survival density and then represent the barrier option price as the one-dimensional integral of its payoff function and the survival density. The approximation error of the survival density is also analyzed.

Since even for the SABR model with constant coefficients the closed form expression for the barrier option price is not available (without approximations), various asymptotic methods have also been proposed assuming that the model contains a small parameter, see e.g., \citep{BargerLorig2017, Kato2013} among others.

Alternatively, numerical methods have been used extensively to price options under the SABR model especially when those options have high maturities and, hence, the asymptotic solution of \citep{hagan2002} (for the European option) becomes inaccurate. For instance, in \citep{Thakoor2019} a computational method based on a spectral discretization of the pricing equation is developed for pricing options with discrete barriers under the arbitrage-free SABR model. The high accuracy of the method is established by comparison with special cases of the SABR model where analytical solutions are available.  But basically, in the literature various flavors of Monte Carlo and finite-difference (FD) methods were used to price barrier options under the SABR model with both time-dependent and constant coefficients.

As mentioned in \citep{Hagan2020}, the SABR model is effective at managing volatility smiles, volatility as a function of the strike $K$ at a single expiry date $T$. For each $T$, the SABR parameters $\alpha, \rho, \nu$ are calibrated so that the model's implied volatility curve $\sigma_I\left(T, K\right)$ matches the market's implied volatilities for that $T$. However,
the SABR model is less useful for managing volatility surfaces, volatility as a function of the strike $K$ at multiple exercise dates $T$, which are needed for handling many exotics. Managing volatility surfaces requires a richer model, such as the dynamic SABR model where all the model coefficients are functions of time.

Therefore, in this paper our main interest is pricing barrier options under the time-dependent SABR model where the barrier level could also be a deterministic function of time. We propose a slightly modified version of the SABR model with the following changes: a) for the instantaneous stochastic volatility we assume a lognormal vol-of-vol (as in the original SABR model) but also add a mean-reverting drift; b) this drift has zero mean-reversion level and is linear in $\sigma_t$. The form of the mean-reversion term for $\sigma_t$ is chosen by a tractability argument.  It is inspired by the $\lambda$-SABR model of \citep{Labordere2005}. The assumption about the linear drift can be relaxed, for instance a quadratic drift can be used inspired by the so-called "3/2" stochastic volatility model, \citep{Platen1997,CarrSun}\footnote{In contrast to these papers all coefficients of our model are arbitrary functions of time (in \citep{CarrSun} the mean reversion rate is an arbitrary function of time and the mean-reversion level, vol-of-vol and correlation are constant)}. However, as this will be seen in section~\ref{intScheme} the linear drift allows reduction of 2D integrals to 1D ones, so the corresponding LMVF equation becomes easier to solve.

With all these preliminaries in mind, we define the  model by the following stochastic differential equations (SDEs)\footnote{Note, that the SABR model has been employed for Fixed Income, FX and Equities. Here without loss of generality we formulate it for the Equity world, while, e.g. for FX this setting can be found in \citep{Oosterlee2015}.}
\begin{align} \label{sabr}
dF_t &= \sigma_t F_t^{\beta+1} dW^{(1)}_t \\
d \sigma_t &= - \kappa(t) \sigma_t dt + \gamma(t) \sigma_t dW^{(2)}_t, \nonumber \\
d \langle W^{(1)}_t,  W^{(2)}_t \rangle &= \rho(t) dt, \quad [F,t] \in [0,\infty) \times [0,\infty),
\quad F_0 = F, \quad \sigma_0 = \sigma, \nonumber
\end{align}
\noindent where $t \geq 0$ is the time, $F_{t}$ is the stochastic forward price, $\sigma_t$ is the stochastic volatility, $\kappa(t)$ is the rate of mean-reversion, $\gamma(t)(t)$ is the volatility of volatility (vol-of-vol), $\theta(t)$ is the mean-reversion level (the long-term run), $W^{(1)}$ and $W^{(2)}$ are two standard correlated Brownian motions with the correlation coefficient $\rho(t) \in [-1,1]$, and $\beta$ is the elasticity parameter such that $-1 < \beta<1, \ \beta \neq\{0,-1\}$\footnote{In case $\beta = 0$ this model is the Black-Scholes model, while for $\beta = -1$ this is the Bachelier, or time-dependent Ornstein-Uhlenbeck (OU) model.}. We assume that all parameters of the model are known either as some continuous functions of time $t \in[0, \infty)$, or as a discrete set of $N$ values for some moments $t_{i}, i=1, \ldots, N$.

It is known that for the process $F_t$ to have a unique solution, an explicit boundary condition has to be set
for $-1 <\beta < -1/2$ which is an absorbing boundary condition for the price process to be a martingale
and arbitrage-free. The same boundary condition also holds naturally for $-1/2 \leq \beta < 0$. In this case the SABR
model has a probability mass at the origin for $-1 < \beta < 0$ similar to that of the CEV model. In the $\sigma$ space it is relatively easy to show that for our SV model the boundary $\sigma_t = 0$ is an attainable regular boundary by Feller's classification, \citep{Lipton2001}. Therefore, similar to \citep{GorovoiLinetsky} we always make regular boundaries instantaneously reflecting, and include  regular reflecting boundaries into the state space.  We also assume that infinite boundaries are unattainable.

To attack the problem of pricing barrier options under this model we develop a new method which is an extension of the Generalized integral transform (GIT) method proposed in a series of the authors' papers (some in cooperation with Peter Carr and Alex Lipton) and covered in detail in a recent book \citep{ItkinLiptonMuraveyBook}. This method was originally developed to solve similar problems for various time dependent one-factor models with time-dependent barriers and provide semi-analytical prices of single and double barrier options. Then in \citep{CarrItkinMuravey2021} for the first time this approach was extended to stochastic volatility models. The authors developed the GIT method for pricing barrier options in the time-dependent Heston model where the option price is expressed in a semi-analytical form as a two-dimensional integral. This integral depends on yet unknown function $\Psi(t,v), \ v=\sigma^2$ which is the gradient of the solution at the moving boundary $F = L(t)$ and solves a linear mixed Volterra-Fredholm (LMVF) equation of the second kind also derived in that paper. Numerical examples illustrate high speed and accuracy of the method as compared with the finite-difference approach. In this paper we extend this idea and apply it to the time-dependent SABR model with mean-reversion.

However, here we do it in a slightly different way. We represent the solution of our problem as a weighted sum of the Bessel functions $J_{|\nu|}(x)$, so actually this representation is a Fourier-Bessel series. The corresponding weights  solve a LMVF equation of the second kind which is also derived in the paper.  Thus, instead of the LMVF for $\Psi(t,v)$ we derive a LMVF equation for the transform image $\baru$. Once it is solved, the weights of the Fourier-Bessel expansion become explicitly known, and so  the solution is expressed in closed form with no further integration. Also, in this way we are able to keep time dependence of the model coefficients with no approximation, while if we use the previous version of the GIT method, for this problem it should be much harder if ever possible. Thus, this version of the GIT method is another new result of this paper.

As shown in Section~\ref{Discus} (see also \citep{ItkinLiptonMuraveyBook}), our methods can be used for any sort of barrier options. In this paper as an example we consider just an Up-and-Out barrier Call option with $F_t \in [0,H(t)], \ H(t) > 0$ is the upper barrier. Once $F_t$ hits the barrier, the contract is terminated and the option expires worthless with no rebate paid either at hit or at maturity, i.e.
\begin{equation} \label{bcH}
C(t,H(t),\sigma) = 0,
\end{equation}
\noindent where $C(t,F,\sigma)$ is the option price. This assumption is not restrictive and can be relaxed, \citep{ItkinMuraveyDBFMF}.

Note, that the barrier $H(t)$ is defined as a level of the forward price. However, since it is a function of time, we can set the barrier for the spot price $S_t$, so $S_t \in [H_S(t), \infty$ as well. In this case we have
\begin{equation*}
H(t) = H_S(t) \exp \left[ \int_0^t (r(k) - q(k)) dk \right],
\end{equation*}
\noindent where $r(t), q(t)$ are the deterministic interest rate and continuous dividend. In a similar way the model can be transformed to a similar model with respect to the spot price $S_t$ by introducing a maturity dependent strike $K_S = K \exp \left[ \int_0^T (r(k) - q(k)) dk \right]$.

At the other end of the domain we have $S = 0$ and the Call price vanishes
\begin{equation} \label{bc0}
C(t,0,\sigma) = 0.
\end{equation}
If the process $F_t$ survives till $t=T$, the option holder gets the Call option payoff
\begin{equation} \label{tc}
C(T,F,\sigma) = (F - K)^+.
\end{equation}
\noindent The \eqref{tc} is the terminal condition for our problem. We also assume that $H(T) \ge  K$.

The rest of the paper is organized as follows. In Section~\ref{pricePDE} we consider the PDE for the price of an Up-and-Out barrier Call option and solve it assuming $\rho(t) = 0$ by using the GIT method similar to that developed in \citep{CarrItkinMuravey2020, CarrItkinMuravey2021}.  In Section~\ref{intScheme} we discuss how the LMVF equation derived in Section~\ref{pricePDE} can be solved numerically. In doing so we use the Radial Basis Functions (RBF) method and show that using Gaussian RBFs  makes the problem tractable by reducing 2D integrals in the LVMF equation to the 1D ones. Section~\ref{rhoNonZero} extends our method to the case when $\rho(t) \neq 0$. In Section~\ref{numExp} we describe some numerical experiments where the barrier option prices obtained by using our method and a FD approach are compared. We show that the speed and accuracy of our method are comparable with those of the finite-difference approach at small maturities and outperform them at high maturities even by using a simplistic implementation of the RBF method.

We show that our method outperforms the FD one in both accuracy and speed. Section~\ref{Discus} concludes.

\section{The pricing PDE and its solution} \label{pricePDE}

The Feynman-Kac theorem, \citep{Shreve:1992}, implies that under the risk neutral measure the Call option price $C(t,F,\sigma)$ solves the partial differential equation (PDE)
\begin{align} \label{PDE1}
\fp{C}{t} &+ \dfrac{1}{2} \sigma^2 F^{2(\beta+1)} \sop{C}{F}  +  \dfrac{1}{2} \gamma^2(t) \sigma^2 \sop{C}{\sigma} - \kappa(t) \sigma \fp{C}{\sigma}  + \rho(t) \gamma(t) \sigma^2 F^{\beta+1} \cp{C}{F}{\sigma} = r(t) C,
\end{align}
\noindent subject to the terminal condition in \eqref{tc} and the boundary conditions in \eqref{bc0}. \eqref{bcH}. Similar to \citep{CarrItkinMuravey2020}, by making a change of variables
\begin{equation} \label{tr1}
F = \left(-x \beta \right)^{-1/\beta}, \qquad C(t,F,\sigma) \to u(t,x, \sigma) e^{\int_T^t r(k) d k},
\end{equation}
\noindent we reduce this PDE to the form
\begin{align} \label{PDEx}
\fp{u}{t} &+ \sigma^2 \left[ \dfrac{1}{2} \sop{u}{x} + \frac{b}{x} \fp{u}{x} +  \dfrac{1}{2} \gamma^2(t) \sop{u}{\sigma} + \rho(t) \gamma(t) \cp{u}{x}{\sigma} \right] - \kappa(t) \sigma \fp{u}{\sigma} = 0, \quad b = \frac{\beta + 1}{2\beta}.
\end{align}
When $\sigma$ is not stochastic \eqref{PDEx} is the PDE associated with the one-dimensional Bessel process, \citep{RevuzYor1999}
\begin{equation} \label{BesProc}
d X_t = d W_t  + \frac{b}{X_t} dt.
\end{equation}

As mentioned in \citep{CarrItkinMuravey2020}, to set the boundary and terminal conditions in the new variables, we must distinguish two cases, which are determined by the sign of $\beta$. If $-1 < \beta < 0$, the variable $x$ is defined at
$x \in [0, y(t)]$, with
\begin{equation} \label{ytau1}
y(t) = -\frac{1}{\beta} H^{-\beta}(t) > 0.
\end{equation}
Therefore, the boundary conditions now read
\begin{equation} \label{bc1}
u(t, 0, \sigma) = u(t,y(t), \sigma) = 0,
\end{equation}
\noindent and the terminal condition in \eqref{tc} takes the form
\begin{equation} \label{tc1}
u(T, x, \sigma) = \left[\left(- \beta  x\right)^{-1/\beta } - K\right]^+,
\end{equation}

However, if $0 < \beta < 1$, the left boundary goes to $-\infty$. Therefore, in this case it is convenient to redefine $x \to \bar{x} = -x$. Then $\bar{x}$ is defined at $\bar{x} \in [\bar{y}(t), \infty)$ where
\begin{equation} \label{ytau12}
\bar{y}(t) = \frac{1}{\beta} H^{-\beta}(t) > 0.
\end{equation}
Accordingly, the terminal condition transforms to
\begin{equation} \label{tc2}
u(T,\bar{x}, \sigma) =  \left[\left(\beta  \bar{x}\right)^{-1/\beta } - K\right]^+,
\end{equation}
\noindent and the boundary conditions read
\begin{equation} \label{bc2}
u(t, \bar{x}, \sigma)\Big|_{\bar{x} \to \infty} = u(t,\bar{y}(t), \sigma) = 0.
\end{equation}
In other words, an Up-and-Out Call options written on the underlying process $x$ in new variables behaves like a  Down-and-Out Put option written on the scaled underlying process $\bar{x}$ as in \eqref{ytau12}.  It can also be checked that when $0 < \beta < 1$ the PDE in \eqref{PDEx} remains same but now in the $\bar{x}$ variable.

\subsection{The GIT method for $\rho=0$} \label{rho0sec}

In this section we consider only the uncorrelated case when $\rho=0$. For the sake of concreteness let us assume that $-1 < \beta < 0$. The other case $0 < \beta < 1$ can be treated in a similar way, \citep{CarrItkinMuravey2020}. Following the approach of \citep{CarrItkinMuravey2020} for the CEV process with $x \in [0, y(t)]$ we introduce the following integral transform
\begin{equation} \label{GITdef}
\baru(t,\sigma,p) = \int_0^{y(t)} u(t,x,\sigma) x^{\nu + 1} \Jnu(x p) dx,
\end{equation}
\noindent where $p = a + i\omega$ is a complex number with $-\pi/4 < \arg(\sqrt{p}) < \pi/4$ and $\nu = 1/(2\beta) < 0$, since $\beta < 0$. Multiplying both parts of \eqref{PDEx} by $x^{\nu + 1}\Jnu(x p)$ and integrating on $x$ from $0$ to $y(t)$ we obtain
\begin{align} \label{tr2}
\fp{\bar{u}}{t} &- y'(t) [y(t)]^{\nu + 1} \Jnu(y(t) p) u(t, y(t),\sigma)  + \sigma^2 \left[ J_1 + J_2 +  \frac{1}{2} \gamma^2(t) \sop{\bar{u}}{\sigma} \right] - \kappa(t) \sigma \fp{\bar{u}}{\sigma}
 = 0.
\end{align}
Here
\begin{align} \label{right}
2 J_1 &= \int_0^{y(t)}  x^{\nu+1} \Jnu(x p) \sop{u}{x} dx =  x^{\nu+1} \Jnu(x p) \fp{u}{x} \Bigg|_0^{y(t)} - u(t,x,\sigma)  \fp{}{x} \left(x^{\nu+1}\Jnu(x p) \right)\Bigg|_0^{y(t)} \\
&+ \int_0^{y(t)} u(t,x,\sigma) \sop{}{x} \left(x^{\nu+1} \Jnu(x p) \right) dx, \nonumber \\
J_2 &= \int_0^{y(t)}  x^{\nu+1} \Jnu(x p) \frac{b}{x} \fp{u}{x}  dx =
b\, x^{\nu}  \Jnu(x p) u(t,x, \sigma)\Bigg|_0^{y(t)}  - b \int_0^{y(t)}  u(t,x, \sigma)  \fp{}{x} \left(x^\nu \Jnu(x p) \right) dx, \nonumber
\end{align}
With allowance for the boundary conditions \eqref{bc1}, the sum $J_1 + J_2$ can be expressed as
\begin{align} \label{sumJ}
J_1 + J_2 = y^{\nu+1}(t) \Jnu(y(t) p) \Psi(t, \sigma) &+ \frac{1}{2} \int_0^{y(t)} u(t,x, \sigma) x \left[\frac{1-2 \nu}{x} \fp{}{x} \left(x^\nu \Jnu(x p) \right) + \sop{}{x}\left(x^\nu \Jnu(x p) \right) \right] dx, \nonumber \\
\Psi(t) &= \frac{1}{2}\fp{u}{x}\Big|_{x = y(t)},
\end{align}
\noindent where we took into account that $b = \nu + 1/2$.

Since the function $X(x) = x^\nu \Jnu(x p)$ solves the ordinary differential equation (ODE), \citep{bateman1953higher}
\begin{equation} \label{Bessel_def}
\frac{d^2 X}{dx^2} + \frac{1-2\nu}{x} \frac{d X}{dx}  + p^2 X = 0,
\end{equation}
\noindent we can re-write \eqref{tr2} in the form
\begin{align} \label{PDEv}
\fp{\bar{u}}{t} &+ \sigma^2 \left[ \frac{1}{2} \gamma^2(t) \sop{\bar{u}}{\sigma} - \frac{1}{2} p^2 \bar{u} \right] - \kappa(t) \sigma \fp{\bar{u}}{\sigma} + \sigma^2 y^{\nu+1}(t) \Jnu(y(t) p) \Psi(t, \sigma) = 0.
\end{align}

The terminal condition for \eqref{PDEv} follows from the definitions in \eqref{GITdef} and \eqref{tc1}
\begin{align} \label{tc1bar}
\baru(T, p) =  -\frac{1}{p} \Bigg\{ & (-\beta)^{-1/\beta}
\left[ a^{1-\nu} J_{1-\nu }( a p) - y^{1-\nu}(T) J_{1-\nu }(p y(T))\right]  \\
 &- K \left[ y^{1+\nu}(T) J_{-1-\nu}(p y(T)) - a^{1+\nu} J_{-1-\nu}(a p)\right] \Bigg\}, \qquad
a = -\frac{K^{-\beta}}{\beta}. \nonumber
\end{align}
Note, that since the option payoff doesn't depend on $\sigma$, the terminal condition in \eqref{tc1bar} either doesn't depend on it.

The \eqref{PDEv} is an inhomogeneous PDE in variables $(t,\sigma)$. In contrast to the analogous PDE for the Heston model described and solved in \citep{CarrItkinMuravey2021}, this equation is not affine.

\subsection{Solution of \eqref{PDEv}}

Our approach to solving \eqref{PDEv} consists in few steps.

\paragraph{Step 1.} We begin with making a change of variables
\begin{align} \label{trSig1}
\baru(t,\sigma,p) &= w(\tau,z,p), \quad z(t, \sigma) = \log \sigma + g(t), \quad
\tau(t) = \frac{1}{2} \int_t^T \gamma^2(k) dk, \quad g(t) = \int_T^t  \kappa(k) dk \, - \tau(t).
\end{align}
In new variables \eqref{PDEv} takes the form
\begin{align} \label{PDEv1}
\fp{w}{\tau} &= \sop{w}{z}  + \eta(t,p) e^{2 z} w + \frac{2}{\gamma^2(t)} e^{2(z  - g(t))} y^{\nu+1}(t) \Jnu(y(t) p) \Psi(t, z), \\
t &= t(\tau), \quad \eta(t,p) = - \frac{p^2}{\gamma^2(t)} e^{-2 g(t)}. \nonumber
\end{align}
Surprisingly, the homogeneous version of this PDE has been already considered in \citep{ItkinLiptonMuravey2020, ItkinLiptonMuraveyBook}\footnote{In \citep{ItkinLiptonMuravey2020} we used the quadratic drift, so instead of the term $n_1(t) z$ it gave rise to $n_1(t) e^z$.}. We have mentioned there that by the change of variables $ z \to -z$ and $\tau \to -\iu \tau$, this PDE  transforms into the time-dependent Schr\"{o}dinger equation with the unsteady Morse potential.

By using the Duhamel's principle, one can deduce that the function $w$ solves the LMVF equation of the second kind, \citep{ItkinLiptonMuravey2020}
\begin{align} \label{IntegralEqZCB}
w(\tau,z, p) &= \frac{1}{2\sqrt{\pi}} \Bigg\{ \int_{-\infty}^{\infty} \frac{w(0, \xi, p)}{\sqrt{\tau}} e^{-\frac{(z - \xi)^2}{4\tau}} d\xi +  \int_{-\infty}^{\infty} d\xi \int_0^{\tau} \frac{w(k,\xi,p)}{\sqrt{\tau - k}} e^{-\frac{(z - \xi)^2}{4(\tau - k)}} \eta(t(k),p) e^{2\xi}  dk \nonumber \\
&+  2\int_{-\infty}^{\infty} d\xi \int_0^{\tau} \frac{e^{-\frac{(z - \xi)^2}{4(\tau - k)}}}{\gamma^2(k) \sqrt{\tau - k}}  e^{2(\xi- g(k))} y^{\nu+1}(\tau) \Jnu(y(k) p)\Psi(t(k),\xi) dk \Bigg\},
\end{align}
\noindent where $w(0, z, p)$ is the initial condition. Since $\baru(T,\sigma, p) = \baru(T,p)$ doesn't depend on $\sigma$ we must have
\begin{equation} \label{tcW}
w(0, z,p)  =  \baru(T,p).
\end{equation}

\paragraph{Step 2.}  The inverse transform for \eqref{GITdef} has been constructed in \citep{CarrItkinMuravey2020}. The solution is represented in the form
\begin{equation} \label{invTrStrip}
u(t, x, \sigma) = x^{-\nu} \sum_{n = 1}^{\infty} \alpha_n(t, \sigma) J_{|\nu|} \left( \frac{\mu_n x}{y(t)}\right),
\end{equation}
\noindent where $\mu_n$ is an ordered sequence of the positive zeros of $\Jnu(\mu)$:
\[ J_{|\nu|}(\mu_n) = J_{|\nu|}(\mu_m) = 0, \quad \mu_n > \mu_m > 0, \quad n >m. \]
By definition, this form automatically respects the vanishing boundary conditions for $u(t, x, \sigma)$ at $x \in [0, y(t)]$. We assume that under some mild conditions on coefficients $\alpha_n(t, \sigma)$ this series converges absolutely and uniformly $\forall x \in  [0,y(t)]$ for any $t > 0$.

The functions $\Jnu(\mu_n x), \ n=1,\ldots$ form an orthogonal basis in the space $C[0,1]$ with the scalar product being
\begin{equation} \label{scal}
\langle \Jnu(\mu_k x), \Jnu(\mu_l x) \rangle =  2\int_0^1 \frac{x \Jnu(\mu_k x) \Jnu(\mu_l x) dx}{J_{|\nu| +1} (\mu_k)  J_{|\nu| + 1}(\mu_l)}  =
\begin{cases}
 1, \quad k = l, \\
 0, \quad k \neq l.
 \end{cases}
 \end{equation}

Therefore, applying the direct integral transform in \eqref{GITdef} to both parts of \eqref{invTrStrip} we obtain
\begin{equation}
\bar{u}(t, \sigma, p) =  y^2(t) \sum_{n = 1}^{\infty} \alpha_n(t, \sigma) \int_0^{1} \hx   J_{|\nu|} \left( \mu_n \hx\right) \Jnu(p y(t) \hx) d\hx, \qquad \hx = x/y(t).
\end{equation}
With allowance for \eqref{scal}, this equation can be solved for each coefficient $\alpha_n(t,\sigma)$ to yield
\begin{equation} \label{alphaCoef}
\alpha_n(t,\sigma) = 2 \frac{\bar{u}\left(t, \sigma, \mu_n /y(t)\right)}{y^2(t) J_{|\nu|+ 1}^2 (\mu_n)}.
\end{equation}

\paragraph{Step 3.} Using the inverse transform in \eqref{invTrStrip} and coefficients $\alpha_n(t,\sigma)$ found in \eqref{alphaCoef}, the solution $u(t,x, \sigma$ can be represented as
\begin{equation} \label{finSol}
u(t, x, \sigma) = 2 x^{-\nu} \sum_{n = 1}^{\infty} \frac{\bar{u}\left(t, \sigma, \mu_n /y(t)\right)}{y^2(t) J_{|\nu| + 1}^2 (\mu_n)} J_{|\nu|} \left( \frac{\mu_n x}{y(t)}\right).
\end{equation}
Differentiating both parts of this representation by $x$, setting $x = y(t)$  and using the identity
$J_{|\nu| - 1} (\mu_n) = - J_{|\nu| + 1}(\mu_n)$, \citep{as64}, yields
\begin{equation} \label{phiBaru}
\Psi(t,\sigma) = -\frac{1}{y^{\nu+3}(t)} \sum_{n = 1}^{\infty} \frac{\mu_n } { J_{|\nu| + 1} (\mu_n)} \bar{u}\left(t, \sigma, \mu_n /y(t)\right).
\end{equation}
Finally, substituting this expression into \eqref{IntegralEqZCB} we obtain a LMVF equation of the second kind for $w(\tau,z,p)$
\begin{align} \label{VoltBaru}
w(\tau,z, p) &= \bar u(T,p)  +  \frac{1}{2\sqrt{\pi}} \int_{-\infty}^{\infty} d\xi \int_0^{\tau} \frac{w(k,\xi,p)}{\sqrt{\tau - k}} e^{-\frac{(z - \xi)^2}{4(\tau - k)}} \eta(k,p) e^{2\xi}  dk \\
&-  \frac{1}{\sqrt{\pi}} \sum_{n = 1}^{\infty} \frac{\mu_n} {J_{|\nu| + 1}(\mu_n)} \int_{-\infty}^{\infty} d\xi \int_0^{\tau} \frac{y^{\nu + 1}(\tau) w\left(k, \xi, \mu_n / y(k))\right)}{y^{\nu+ 3}(k) \gamma^2(k) \sqrt{\tau - k}}
\Jnu(y(k) p) e^{-\frac{(z - \xi)^2}{4(\tau - k)} + 2(\xi - g(k)) } dk. \nonumber
\end{align}
This equation has to be solved for each $p = p_i \equiv \mu_i/y(\tau), \ n=1,2,\ldots$. Once this is done, the solution of the whole problem is given be \eqref{trSig1}. In particular, we are interested in $\tau = \tau(0) = \frac{1}{2} \int_0^T \gamma^2(k) dk$.

It is easy to see that if $y(t) = y -  const$ the last term in \eqref{VoltBaru} vanishes, and thus for each $w(\tau,z, \mu_i/y(\tau))$ we obtain an independent LMVF equation
\begin{align} \label{VoltBaru3}
w(\tau,z, \mu_i/y) &= \bar u(T,\mu_i/y) -  \frac{\mu^2_i}{2\sqrt{\pi} y^2} \int_{-\infty}^{\infty} d\xi \int_0^{\tau} \frac{w(k,\xi,\mu_i/y)}{\gamma^2(k) \sqrt{\tau - k}} e^{-\frac{(z - \xi)^2}{4(\tau - k)} + 2 (\xi-g(k))}  dk, \quad i=1,2,\ldots.
\end{align}
Once they are solved, the solution of the whole problem is given be \eqref{trSig1}.

Looking at \eqref{VoltBaru} one can discover that actually here we developed  another version of the GIT method which, to the best of the author's knowledge, is not known in the literature. Indeed, the first versions of the GIT method have been proposed in \citep{kartashov2001} (see also reference therein), and as applied to mathematical finance in \citep{CarrItkin2020jd}. Then, significant development of this method for various problems and domains was done in a series of the author's papers in collaboration with Peter Carr and Alex Lipton, see \citep{ItkinLiptonMuraveyBook}. In all cases the core of the method is to obtain a Volterra equation of the second kind for the gradient $\Psi(t)$ (in the one-dimensional case), or the LMVF equation for $\Psi(t,\sigma)$ (in the two dimensional case discussed in \citep{CarrItkinMuravey2021}). Once it is solved, the solution is expressed in a semi-analytical form via an integral  of $\Psi$.

Here we do it in a slightly different way. Instead of the LMVF equation for $\Psi$ we derive a LMVF equation for the image $\baru$. Once it is solved, the solution is expressed in closed form (with no further integration). Also, in this way we are able to keep time dependence of the coefficients $\kappa(t), \gamma(t)$, while if we use the previous version of the GIT method for solving this problem under consideration, this should be much harder if ever possible.

\subsection{Analysis of convergence} \label{converge}

\begin{figure}[!htb]
\begin{center}
\hspace*{-0.3in}
\fbox{\includegraphics[width=0.7\textwidth]{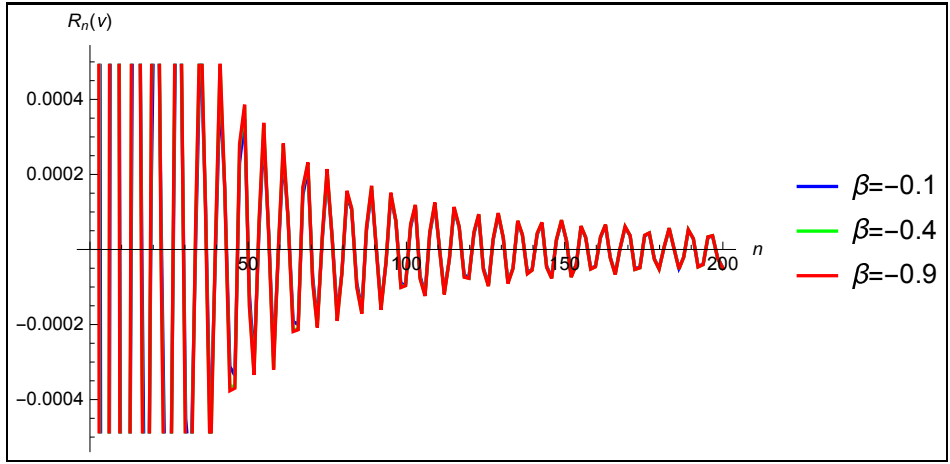}}
\end{center}
\caption{The ratio $R_n(\nu) = \frac{J_{|\nu| + 1}(\mu_n)} {\mu_n}$ as a function of $n$ for $\beta = -0.1, -0.4, -0.9$.}
\label{ratio}
\end{figure}

Recall, that $\nu = 1/(2\beta)$ and we consider the case $-1 < \beta < 0$. A simple analysis shows that the ratio $\frac{\mu_n} {J_{|\nu| + 1}(\mu_n)}$ rapidly grows with the increase of $n$. To illustrate, in Fig.~\ref{ratio} the inverse ratio  $R_n(\nu) = \frac{J_{|\nu| + 1}(\mu_n)} {\mu_n}$ is depicted as a function of $n$ for $\beta = -0.1, -0.4, -0.9$.

To compensate this growth of the coefficients the solution $w(\tau,z, \mu_n/y(\tau))$ should decrease faster than the coefficients to converge. However, we don't know in advance the speed of this decrease. Therefore, numerical solution of \eqref{VoltBaru} could be unstable.  To get rid of this effect we make a change of the dependent variable $w(\tau,z, \mu_n/y(\tau)) \mapsto \barw(\tau,z, \mu_n/y(\tau)) = w(\tau,z, \mu_n/y(\tau))/R_n(\nu)$. Then \eqref{VoltBaru} transforms to
\begin{align} \label{VoltBaru2}
R_i(\nu) \barw(\tau,z, \mu_i/y(\tau)) &= w(0, \mu_i/y(\tau)) -  \frac{\mu_i^2 R_i(\nu)}{2\sqrt{\pi}} \int_{-\infty}^{\infty} d\xi \int_0^{\tau} \frac{\barw(k,\xi,\mu_i/y(\tau))}{y^2(\tau) \gamma^2(k)\sqrt{\tau - k}} e^{-\frac{(z - \xi)^2}{4(\tau - k)} + 2( \xi-g(k))}  dk  \\
&-  \frac{1}{\sqrt{\pi}} \sum_{n = 1}^{\infty} \int_{-\infty}^{\infty} d\xi \int_0^{\tau} \frac{y^{\nu+ 1}(\tau)\barw\left(k, \xi, \mu_n /y(k)\right)}{\gamma^2(k) y^{\nu + 3}(k) \sqrt{\tau - k}} \Jnu\left(\mu_i \frac{y(k)}{y(\tau)}\right) e^{-\frac{(z - \xi)^2}{4(\tau - k)} + 2(\xi- g(k))} dk. \nonumber
\end{align}

Using \eqref{tcW} and \eqref{tc1bar}, one can find that the first term in the RHS of \eqref{VoltBaru2} has the form
\begin{equation} \label{asym1}
w(0, \mu_i/y(0)) \propto \frac{J_{1-\nu}(\mu_i)}{\mu_i} .
\end{equation}
 With the increase of $i$ the roots $\mu_i \to \infty$. Therefore,  in this limit the function in \eqref{asym1} vanishes. Also, since $\nu = 1/(2\beta) < 0$ functions $w(0, \mu_i/y(0))$ in this limit rapidly tend to zero. Same is true for coefficients $R_i(\nu)$ in the LHS of \eqref{VoltBaru2}. Thus, assuming $\barw(k,\xi,\mu_i/y(k))$ is finite for all $i = 1,2,\ldots$, at high $i$ \eqref{VoltBaru2} asymptotically takes the form
\begin{align} \label{VoltBaru4}
0 &= - \frac{1}{2\sqrt{\pi}}\int_{-\infty}^{\infty} d\xi \int_0^{\tau} dk \frac{e^{-\frac{(z - \xi)^2}{4(\tau - k)} + 2( \xi-g(k))}}{y^2(\tau) \gamma^2(k)\sqrt{\tau - k}} \left[ \barw(k,\xi,\mu_i/y(k)) + 2 \frac{y^{\nu+1}(k)}{y^{\nu+1}(\tau)} \frac{\Jnu\left(\mu_i \frac{y(k)}{y(\tau)}\right)}{J_{|\nu| + 1} (\mu_i) \mu_i} \sum_{n = 1}^{\infty} \barw\left(k, \xi, \mu_n /y(k)\right)\right].
\end{align}
 With the increase of $i$ the term $P_i(\nu) = J_{|\nu| + 1} (\mu_i) \mu_i$ slowly grows, so
 $|J_{|\nu| + 1} (\mu_i) \mu_i| \to \infty, \ i \to \infty$. Therefore, this forces the solution of \eqref{VoltBaru4}
 to vanish, i.e. $\barw(k,\xi,\mu_i/y(k)) \to 0, \, i \to \infty$.

\begin{figure}[!htb]
\begin{center}
\hspace*{-0.3in}
\fbox{\includegraphics[width=0.7\textwidth]{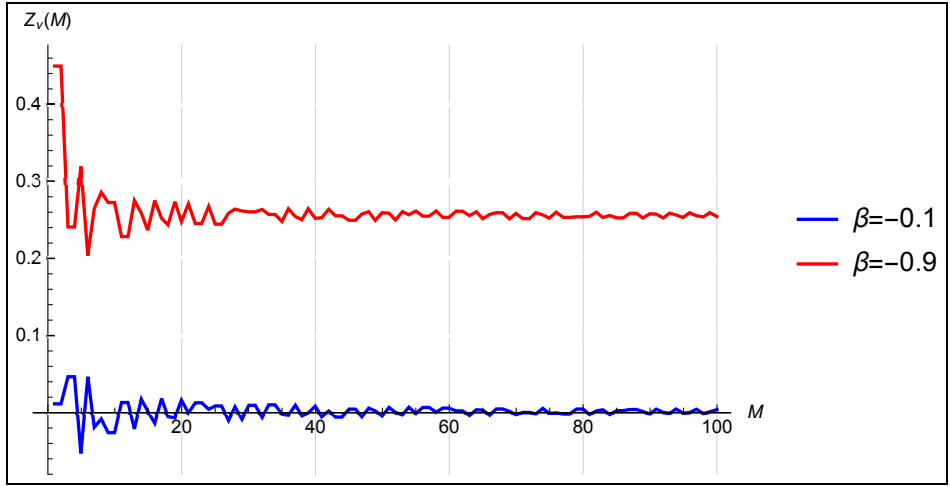}}
\end{center}
\caption{The behavior of $Z_\nu(M)$ as a function of $M$ for $\beta = -0.1, -0.9$ where $\eta = x/y(t), \ 0 \le \eta \le 1$ and $\eta = 0.3$.}
\label{convFig}
\end{figure}

 Due to this behavior of $\barw(k,\xi,\mu_i/y(k))$ and $P_i(\nu)$ at high $i$, the solution in \eqref{finSol}
\begin{align} \label{finSol1}
u(t, x, \sigma) &= 2 x^{-\nu} \sum_{n = 1}^{\infty} \bar{u}\left(t, \sigma, \mu_n /y(t)\right) \frac{1}{y^2(t) J_{|\nu| + 1}^2 (\mu_n)} J_{|\nu|} \left( \frac{\mu_n x}{y(t)}\right) \\
&= \frac{2 x^{-\nu}}{y^2(t)} \sum_{n = 1}^{\infty} \frac{\barw\left(t, \sigma, \mu_n /y(t)\right)  }{J_{|\nu| + 1} (\mu_n) \mu_n} J_{|\nu|} \left( \frac{\mu_n x}{y(t)}\right) \nonumber
\end{align}
\noindent converges.  To illustrate the rate of convergence, in \eqref{finSol1} we drop the dependence of $\barw\left(t, \sigma, \mu_n /y(t)\right)$ on $n$ and plot $Z_\nu(M) = \sum_{n=1}^M J_{|\nu|} \left( \frac{\mu_n x}{y(t)}\right)/[J_{|\nu| + 1} (\mu_n) \mu_n]$ as a function of $M$  for $\beta = -0.1, -0.9$., where  we use $\eta = x/y(t), \ 0 \le \eta \le 1$ and $\eta = 0.3$. It can be seen in Fig.~\ref{convFig} that this function becomes almost flat at $M > 60$.

This behavior can be made even more transparent if we take into account, that at $n \to \infty$
\begin{align}
\mu_{n} &\sim  n \pi + \frac{\pi}{2} \left(|\nu| - \frac{1}{2}\right) - \frac{4 |\nu|^{2} - 1}{8 \pi (n + 1/2(|\nu| - 1/ 2))} + \mathcal{O}\left(\frac{1}{n^{3}}\right), \\
J_{|\nu|} (\mu_n) &\sim \left(\frac{2}{\pi \mu_n}\right)^{\frac{1}{2}}
\left(\cos \omega \sum_{k=0}^{\infty}(-1)^{k} \frac{a_{2 k}(|\nu|)}{\mu_n^{2 k}}
- \sin \omega \sum_{k=0}^{\infty}(-1)^{k} \frac{a_{2 k+1}(|\nu|)}{\mu_n^{2 k+1}}\right), \nonumber \\
a_{k}(|\nu|) &= \frac{\left(4 |\nu|^{2}-1^{2}\right)\left(4 |\nu|^{2}-3^{2}\right) \cdots\left(4 |\nu|^{2}-(2 k-1)^{2}\right)}{k ! 8^{k}}, \quad \omega = \mu_n - \frac{1}{2} \pi \left(|\nu| + \frac{1}{2}\right), \nonumber
\end{align}
\noindent see McMahon's expansion in \citep{Watson1966} and \citep{as64}. Taking the leading terms yields
\begin{align}
\frac{J_{|\nu|} \left( \frac{\mu_n x}{y(t)}\right)}{J_{|\nu| + 1} (\mu_n) \mu_n} &= \frac{(-1)^{1-n} \cos \left[\frac{1}{4} \pi  (-2 | \nu | +x (2 | \nu | -4 n+1)+1)\right]}{\pi  n \sqrt{x}} + O(n^{-2}) \\
&= - \frac{1}{2 \pi  n \sqrt{x}} \left\{ \cos \left[n \pi (1+x) - q \right] + \cos \left[n \pi (1-x) + q \right] \right\} + O(n^{-2}), \nonumber \\
q &= \frac{1}{4} \pi  (2 (x-1) | \nu | +x+1). \nonumber
\end{align}
Hence, at $n \to \infty$ the function $Z_\nu(M)$ behaves similar to the sum $I_1 + I_2$ of two cosine integrals
\begin{equation}
I_1 = \int_{n_0}^{\infty } \frac{ \cos ( n \pi (1+x) - q)}{n} \, dn, \qquad
I_2 = \int_{n_0}^{\infty } \frac{ \cos ( n \pi (1-x) + q)}{n} \, dn,  \quad x \ne 1,
\end{equation}
\noindent i.e. converges.

Since the function $\barw\left(t, \sigma, \mu_n /y(t)\right)$ also decreases with $n$ growing, the solution in \eqref{finSol1} converges by the Lebesgue's dominated convergence theorem, and even faster than this is shown in Fig.~\ref{convFig}.

Recall, that the representation of the solution $u(t, x, \sigma)$ is given in \eqref{finSol}. One can observe that this semi-analytical solution can be treated as the expansion of $u(t, x, \sigma)$ into Fourier-Bessel series with coefficients proportional to function $\bar{u}\left(t, \sigma, \mu_n /y(t)\right)$. Each such a function solves either the LMVF equation \eqref{VoltBaru3} (if $y(t) = y$ - const), or a system of LMVF equations in \eqref{VoltBaru2} (in a general case).

Alternatively, using the representation, \citep{Watson1966}
\begin{equation}
2 \sum_{n = 1}^{\infty} \frac{J_{|\nu|} \left( \frac{\mu_n x}{y(t)}\right)}{J_{|\nu| + 1} (\mu_n) \mu_n} =
\left(\frac{x}{y(t)}\right)^{-\nu}, \quad \nu < 0, \ x \ne y(t),
\end{equation}
\noindent the solution in \eqref{finSol1} can be bounded from above as
\begin{align} \label{finSol2}
u(t, x, \sigma) &=  \frac{2 x^{-\nu}}{y^2(t)}  \sum_{n = 1}^{\infty} \frac{\barw\left(t, \sigma, \mu_n /y(t)\right)  }{J_{|\nu| + 1} (\mu_n) \mu_n} J_{|\nu|} \left( \frac{\mu_n x}{y(t)}\right)
\le \frac{2 x^{-\nu}}{y^2(t)}  \max_{n}|\barw\left(t, \sigma, \mu_n /y(t)\right)|
\sum_{n = 1}^{\infty} \frac{J_{|\nu|} \left( \frac{\mu_n x}{y(t)}\right)}{J_{|\nu| + 1} (\mu_n) \mu_n}  \\
&= \frac{x^{-2\nu}}{y^{2-\nu}(t)}   \max_{n}|\barw\left(t, \sigma, \mu_n /y(t)\right)|, \qquad x \ne y(t). \nonumber
\end{align}
Since $\barw\left(t, \sigma, \mu_n /y(t)\right)$ is the decreasing function of $n$, and thus finite, the RHS of \eqref{finSol2} is finite. Therefore, by the Lebesgue's dominated convergence theorem the solution of our problem is finite.

\begin{figure}[!htb]
\centering
\hspace*{-0.1in}\subfloat[]{\includegraphics[width=0.53\textwidth]{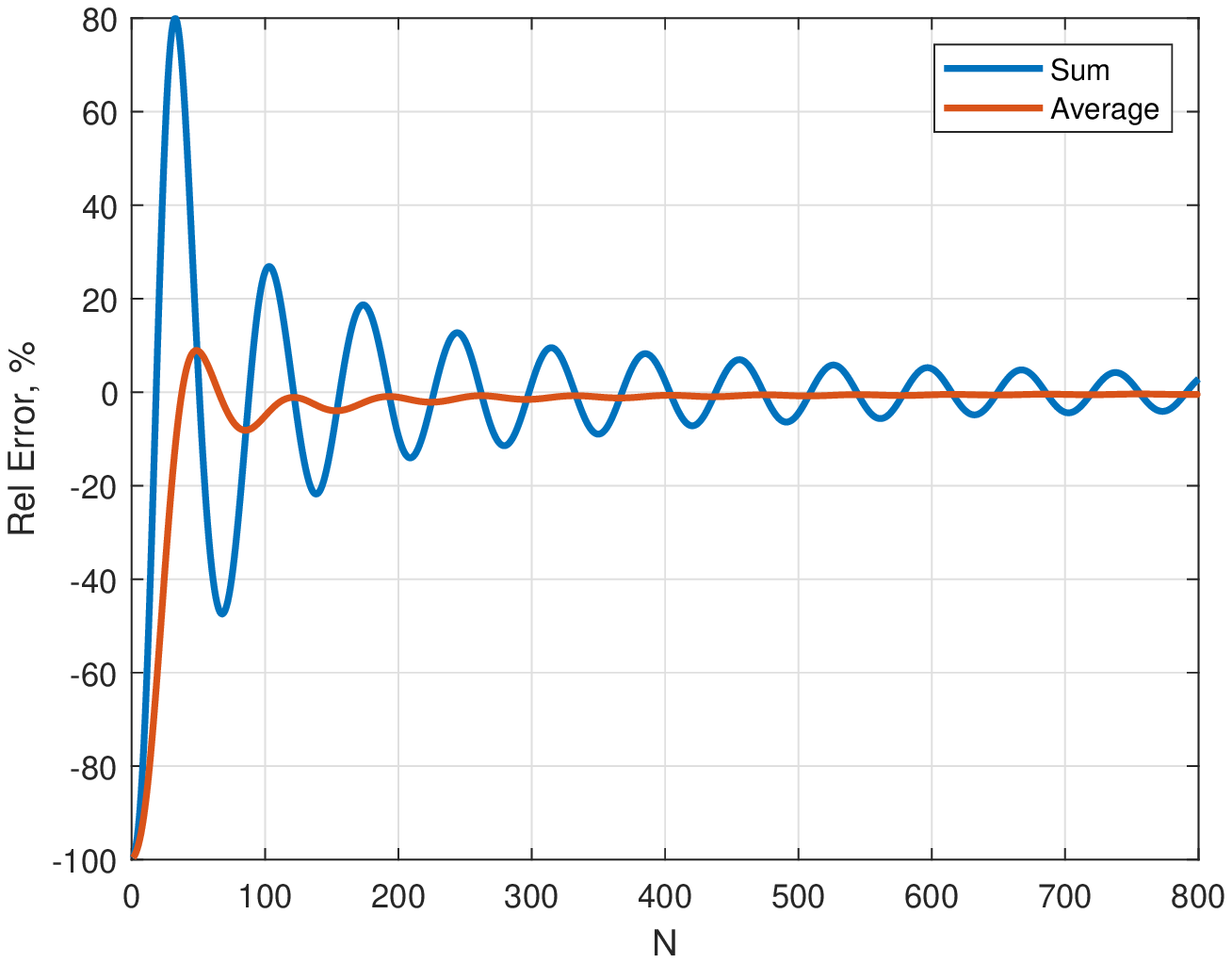}}
\hspace*{-0.2in}\subfloat[]{\includegraphics[width=0.53\textwidth]{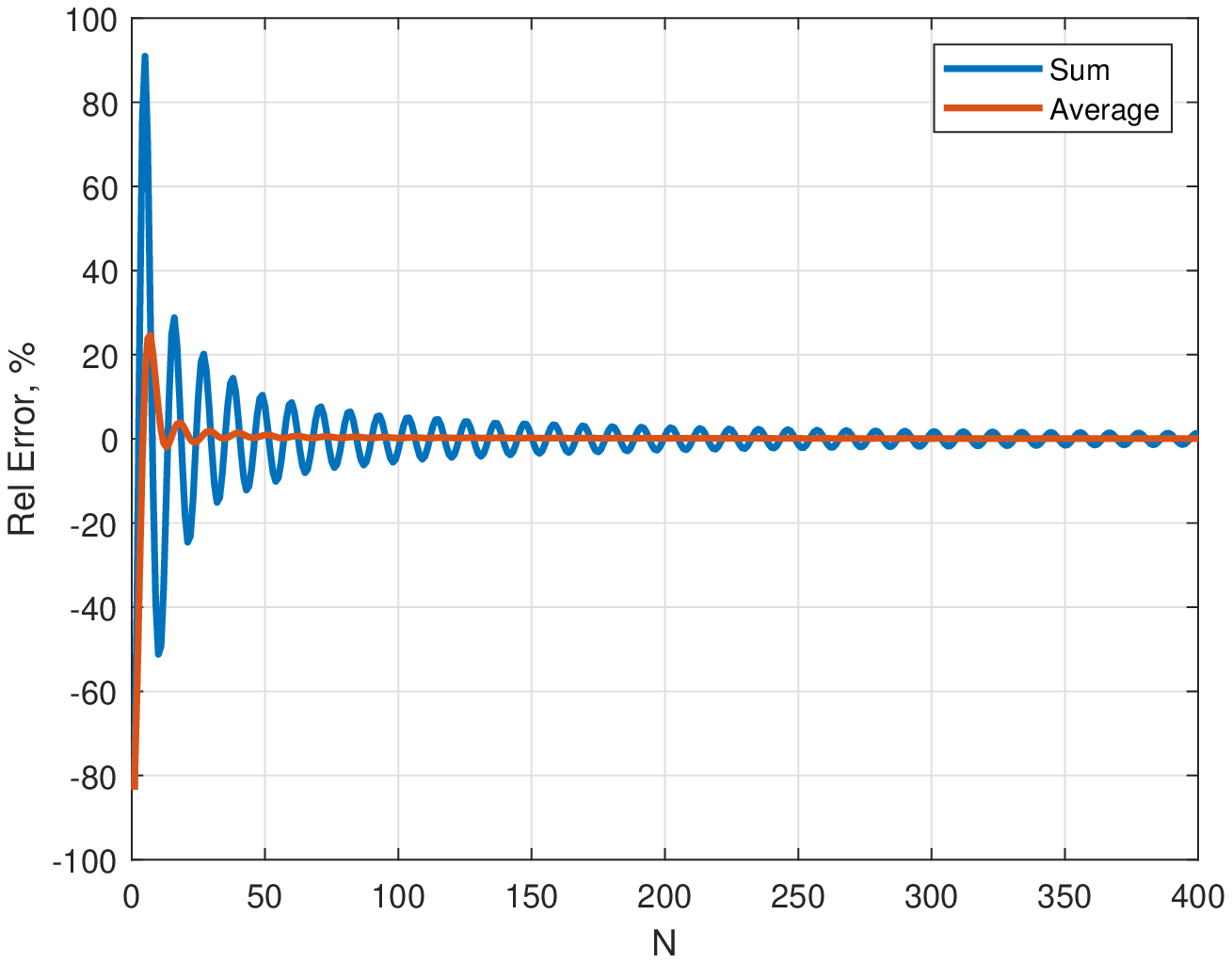}}
\caption{The relative error in \% of the Call option value at maturity as a function of the number of terms $N$ in the Fourier-Bessel series in \eqref{finSol} for (a) $\beta = -0.1$; (b) - $\beta = -0.7$.}
\label{serFig}
\end{figure}

It is, however, well known that Fourier–Bessel series are imposed to a Gibbs Phenomenon. Also, even for continuous functions they  could converge very slow, \citep{Pinsky1992}. In our case this depends on the value of $\beta$. As an example, let us put $\tau=0, \ y(t) = y = const$, so from \eqref{VoltBaru3} we obtain an immediate solution
\begin{equation}
w(0,z, \mu_i/y) = \bar u(T,\mu_i/y).
\end{equation}
Subsisting this solution into \eqref{finSol} we expect to get the option price equal to the option payoff. We truncate
infinite series in \eqref{finSol} up to $N$ terms and compute the option price by summation. Convergence of the series in \eqref{finSol} is depicted in Fig.~\ref{serFig} as a function of $N$ for several $\beta$. Other parameters of the test are: $F = 60, K = 55$.
Fortunately, the Abelian theorem, \citep{Hardy1991},  states that for any summation method $L$ if $u = \{u_n\}$ is a convergent sequence, with limit $U$ at $n \to \infty$, then $L(u) = U$. In our case let us define $L$ to be the arithmetic means of the first $N$ terms of $u$ (the Cesaro method). One can prove that if $u$ does converge to $U$, then so does the sequence $Lu$. The Call option price obtained by using the Cesaro method is also presented in Fig.~\ref{serFig}.  It can be seen that this method significantly improves the convergence, especially at low $|\beta| \ll 1$.

\section{Solution of the LMVF equation \eqref{VoltBaru}} \label{intScheme}

It is known that solving the PDE in \eqref{PDE1} can be done by using traditional numerical methods, for instance the FD approach. As compared with this method, solving the LMVF equation \eqref{VoltBaru2} is a bit less straightforward despite this approach can have some advantages. First, for doing so we can use the RBF method. We will show below that by using Gaussian RBFs the integrals on $\xi$ in \eqref{VoltBaru2} can be taken in closed form. Thus, instead of double integrals, this equation will contain only single ones.

However, first we need to decide how to deal with the last term in \eqref{VoltBaru2} which is new and doesn't appear
for the Heston model which is discussed in \citep{CarrItkinMuravey2021}. The problem with this term lies in the fact that it contains summation over all functions $\barw(k,\xi,\mu_i/y(k)), \ i=1,\ldots,$ and, thus, doesn't allow factorization of \eqref{VoltBaru2} into a system of independent LMVF for each $i \in [1,\infty)$. Therefore, we suggest solving \eqref{VoltBaru2} iteratively.

The main idea behind this approach is as follows.  As was mentioned in Section~\ref{converge}, functions $\barw(k,\xi,\mu_i/y(k))$ tend to zero at high $i$. Therefore, we can truncate the series and solve \eqref{VoltBaru2} only for $i \in [1,M]$, where $M$ is finite and, perhaps, not so large. To begin, let us assume that $W(k,\xi) =
\sum_{i=1}^\infty \barw(k, \xi, \mu_i/y(k)) = 0$. Then \eqref{VoltBaru2} factorizes into $M$ independent LMVF equations (for each $i=[1,M]$) which are given in \eqref{VoltBaru3}.

Suppose this equation is solved numerically and the solution is denoted as $\barw_1(\tau,z, \mu_i/y(\tau))$. Then at the next iteration the last term in \eqref{VoltBaru2} can be replaced with the term
\begin{align}
\Lambda_1(\tau,z) = &-  \frac{1}{\sqrt{\pi}} \sum_{n = 1}^{\infty} \int_{-\infty}^{\infty} d\xi \int_0^{\tau} \frac{ y^{\nu+ 1}(\tau)\barw_1\left(k, \xi, \mu_n /y(k)\right)}{\gamma^2(k) y^{\nu + 3}(k) \sqrt{\tau - k}} \Jnu\left(\mu_i \frac{y(k)}{y(\tau)}\right) e^{-\frac{(z - \xi)^2}{4(\tau - k)} + 2(\xi- g(k)])} dk.
\end{align}
Hence, instead of \eqref{VoltBaru2}, at the second iteration we have to solve $M$ independent LMVF (for each $i \in [1,M]$ of the form
\begin{align} \label{iter1}
\barw_2\left(\tau,z, \frac{\mu_i}{y(\tau)}\right) &- \frac{\mu_i^2}{2\sqrt{\pi}} \int_{-\infty}^{\infty} d\xi \int_0^{\tau} \frac{\barw_2(k,\xi,\frac{\mu_i}{y(\tau)})}{y^2(\tau) \gamma^2(k) \sqrt{\tau - k}} e^{-\frac{(z - \xi)^2}{4(\tau - k)} + 2(\xi-g(k))} dk = \baru\left(T, \frac{\mu_i}{y(0)}\right) + \frac{\Lambda_1(\tau,z)}{R_i(\nu)}.
\end{align}
Once this is done, this process can be continued until convergence to the desired tolerance in reached at some iteration $j$, so the final solution is given by $w_j(\tau,z, \mu_i/y(\tau))$. The complexity of this process is $j M C_{LMVF}$ where $C_{LMVF}$ is the complexity of solving a single LMVF of the type \eqref{iter1}. Since all LMVF are independent, this process is naturally parallel and can be run on a parallel architecture very efficiently.

The LMVF of the type \eqref{iter1} can be solved by using, for instance, the RBF method which we describe in the next section.

\subsection{The RBF method} \label{sec:RBF}

For the description of the RBF method as applied to the solution of integral equations we refer the reader to
\citep{Assari2019,Zhang2014}  and references therein. Here we present only a short exposition of the method.

A function $\Phi: \mathbb{R}^{d} \rightarrow \mathbb{R}$ is called to be radial if there exists a univariate function $\phi:[0, \infty) \rightarrow \mathbb{R}$ such that
\begin{equation}
\Phi(\mathbf{x})=\phi(r)
\end{equation}
\noindent  where $r=\|\mathbf{x}\|$ and $\|\cdot\|$ is some norm in $\mathbb{R}^{d}$. In this paper we consider just the Euclidean norm. Let $\chi=\left\{\mathbf{x}_{1}, \ldots, \mathbf{x}_{N}\right\}$ be a set of scattered points selected in the domain $\Omega \subset \mathbb{R}^{d}$. A function $u(\mathbf{x})$ at an arbitrary point $\mathbf{x} \in \Omega$ can be approximated by using the global radial function $\phi(\|\mathbf{x}\|)$ via a linear combination
\begin{equation} \label{GlInterp}
u(\mathbf{x}) \approx \mathcal{G}_{N} u(\mathbf{x})=\sum_{i=1}^{N} c_{i} \phi\left(\left\|\mathbf{x}-\mathbf{x}_{i}\right\|\right), \quad \mathbf{x} \in \Omega,
\end{equation}
\noindent where the coefficients $\left\{c_{1}, \ldots, c_{N}\right\}$ are determined by the interpolation conditions
\begin{equation}
\mathcal{G}_{N} u\left(\mathbf{x}_{i}\right)=u\left(\mathbf{x}_{i}\right)=u_{i}, \quad i=1, \ldots, N.
\end{equation}

In the literature various choices of the RBFs exist. Among others, here by a tractability argument we prefer to use the Gaussian RBF
\begin{equation} \label{gauss}
\Phi(\mathbf{x}) = e^{-\varepsilon \|\mathbf{x}\|^{2}},
\end{equation}
\noindent where $\varepsilon>0$ is the shape parameter. This function is strictly positive-definite in $\mathbb{R}^{d}$ and, therefore, the expansion in \eqref{GlInterp} is non-singular.

The advantage of using the Gaussian RBFs in our particular case lies in the fact that we can eliminate one numerical integration in \eqref{VoltBaru2}. Suppose we are at the $m$-th iteration of the process described in the previous section. Let us use \eqref{GlInterp} for interpolation of $\barw(\tau,z, \mu_i/y(\tau))$ with the RBF given in \eqref{gauss}
\begin{equation}
\barw(\tau,z,\mu_i/y(\tau)) = \sum_{l}^{N_l}  \sum_{j}^{N_j} c^{(i)}_{(j,l),m} e^{- \varepsilon(\tau-\tau_j)^2 - \varepsilon(z-z_l)^2}.
\end{equation}
Substituting this expression into \eqref{VoltBaru2}, after some algebra we obtain
\begin{align} \label{VoltRBF}
\baru & (T, \mu_i/y(\tau)) = \sum_{l}^{N_l}  \sum_{j}^{N_j} c^{(i)}_{(j,l)} \left[e^{- \varepsilon(\tau-\tau_j)^2 - \varepsilon(z-z_l)^2}  + \mu_i^2  \int_0^\tau \frac{e^{-2 g(k) - \varepsilon(k-\tau_j)^2}}{y^2(\tau) \gamma^2(k)} I_{jl}(z,\tau, k) \, dk \right], \\
I_{jl}(z,\tau, k) &= \frac{1}{2\sqrt{\pi(\tau - k)}} \int_{-\infty}^{\infty} d\xi e^{-\frac{(z - \xi)^2}{4(\tau - k)} - \varepsilon(\xi-z_l)^2 + 2 \xi}
= \frac{1}{\sqrt{h}} \exp\left[-\frac{\epsilon  (z- z_l)^2-2 z -4 (\tau-k) (2 z_l \epsilon +1)}{h}\right],  \nonumber \\
h &= h(\tau,k) = 1 + 4 \varepsilon(\tau-k). \nonumber
\end{align}
Here the term $\Lambda_{m}(\tau,z)$ can be represented as
\begin{align} \label{Lambda_m}
\Lambda_{m}(\tau,z) = &- \frac{1}{\sqrt{\pi}} \sum_{n = 1}^{\infty} \sum_{l}^{N_l}  \sum_{j}^{N_j} c^{(n)}_{(j,l),m} \int_0^{\tau} dk \frac{y^{\nu+1}(\tau)e^{- \varepsilon(k-\tau_j)^2  - 2 g(k)}}{\gamma^2(k) y^{\nu+ 3}(k) \sqrt{\tau - k}} \Jnu\left(\mu_i \frac{y(k)}{y(\tau)}\right) \int_{-\infty}^{\infty} d\xi  e^{-\frac{(z - \xi)^2}{4(\tau - k)} + 2 \xi -  \varepsilon(\xi-z_l)^2}  \nonumber \\
&=  - 2 \sum_{n = 1}^{\infty} \sum_{l}^{N_l}  \sum_{j}^{N_j} c^{(n)}_{(j,l),m} \int_0^{\tau} dk \frac{ y^{\nu+1}(\tau)e^{- \varepsilon(k-\tau_j)^2}}{y^{\nu+ 3}(k) \gamma^2(k)} \Jnu\left(\mu_i \frac{y(k)}{y(\tau)}\right)
I_{jl}(z,\tau, k).
\end{align}

The remaining integrals in $k$ in the RHS of \eqref{VoltRBF} in general cannot be taken in closed form and should be discretized by using some quadrature rule. This is because the model coefficients $\kappa(t), \gamma(t)$ has to be found by calibrating the model to market data. Alternatively, if practitioners are able to guess the functional form of these coefficients, e.g., $\kappa(t) = \kappa_1 - \kappa_2 e^{-\kappa_3 t}, \ \kappa_i = const, \ i=1,2,3$, then they need to calibrate just the constants $\kappa_i, \ i=1,3$. In this case the integral on $k$ in \eqref{VoltRBF} can be taken explicitly for some families of functions, for instance for linear and quadratic functions of $\tau$.

The RBF methods belong to the class of meshfree methods. That means that no regular grid in $\tau_j, z_l$ is required to run it (in contrast, e.g. to the FD method).  Therefore, taking a 2D set of collocation nodes $\{\tau_{j(l)}, z_l\}, \ l = 1,...,\bar{l}, \ j(l)=1,...,\bar{j}(l)$ one can substitute them into \eqref{VoltRBF} and get a system of linear equations for the coefficients $c_{j,l}$. For instance, in case of the regular grid with $N_l$ nodes in $z_l$ and $N_j$ nodes in $\tau_j$ we have $N_j N_l$ unknown coefficients which solve the system of $N_j N_l$ linear equations. The matrix of this system is dense, and therefore complexity of solving this system by using the direct solver is $O(N^3_j N^3_l)$.
Obviously, this result is not satisfactory from computational point of view. Iterative solvers can improve this especially when a suitable preconditioner can be constructed.

It is, however, well-known, e.g., \citep{McCourt2012}, the global Gaussian RBF method leads to a notoriously ill-conditioned interpolation matrix whenever $\varepsilon$ is small and the set of basis functions in \eqref{gauss} becomes numerically linearly dependent on $\mathbb{R}^d$. This leads to severe numerical instabilities and limits the practical use of Gaussians — even though one can approximate a function with the Gaussian kernel with spectral approximation rates. On the other hand, small $\varepsilon$ provide better accuracy, and so have to be considered as an option for an accurate pricing. It is also known that if $\varepsilon$ is kept fixed, convergence stagnates even if $N_j, N_l$ grow, and if $N_j, N_l$ are fixed, the error blows up with the decrease of $\varepsilon$.

One of the ideas to "fix" all these problems is using a "better basis" for RBF interpolation to obtain well-conditioned (and therefore numerically stable) interpolation, see \citep{McCourt2012} and references therein. Among others, let us mention the RBF-QR method, \citep{Fornberg2, Larsson2, Larsson7}. Another approach would be instead of the global RBF method using its localized version. The main improvement comes from the fact that when two collocation nodes lie far away from each other, the exponential function in \eqref{VoltRBF} becomes small. Therefore, only the nodes from the close neighborhood of the given node $(\tau_j, z_l)$ should be taken into account in the summation in  \eqref{VoltRBF} (for instance, ${\cal N}$ closest nodes). This is similar to the FD method where numerical approximation of derivatives is provided using just the closest nodes. In more detail, the localized RBF method as applied to solving Volterra integral equations, is described, e.g. in \citep{Assari2019} (see also references therein).

When the localized version of RBF is used, the system matrix becomes sparse.  For a regular grid it becomes block-sparse with the size of the block equal to $O({\cal N})$ (from ${\cal N}$ at edges to $2{\cal N}-1$ in the middle of the matrix). Thus, the complexity of solving this system drops down to $O({\cal N}^2 N_j, N_l)$.

Having this in mind, in this paper we, however, use just a simple version of the global Gaussian RBF method. This is done for two reasons. First, here we want to illustrate that the proposed method gives reasonable option prices why don't require an immediate very high accuracy. Second, even with this global method the speed of computations is better than that of the FD method. Also, using more sophisticated RBF method for solving the LMVF equation thus improving the speed and accuracy is subject of a separate paper which will be presented elsewhere.

Going back to the global RBF method, we can re-write the interpolation conditions in \eqref{GlInterp} in a matrix form
\begin{equation} \label{matSys}
\|B + A \mu_i^2\| |C| = |\baru| + |\Lambda_m|.
\end{equation}
Here $B$ is the $N_j N_l \times N_j N_l$ matrix $\|e^{- \varepsilon(\tau-\tau_j)^2 - \varepsilon(z-z_l)^2}\|$ - the Gaussian kernel for $\tau \in [\tau_1,\ldots,\tau_{N_j}], \ \tau_1 = 0, \ \tau_{N_j} = \tau(0)$, $z \in [z_1,\ldots,z_{N_l}]$, $A$ is a similar matrix $\| I_j,l\|$, $\baru|$ is the $N_j N_l$ column vector with first $N_j$ components equal to $\baru (T, \mu_i/y(\tau_1))$, next $N_j$ components equal to $\baru(T, \mu_i/y(\tau_2))$, and so on, $|\Lambda_m|$ is the $N_j N_l$ column vector, and $C$ is the unknown $N_j N_l$ column vector of coefficients. The \eqref{matSys} is a linear system of algebraic equation whose solution gives us the coefficients $C$.

It is important to underline that for every $i$, the matrices $A, B$ are same, as well as the matrix $\|B + A \mu_i^2\| $ doesn't change in between the iterations on $m$. Therefore, using a good matrix preconditioner helps to reduce numerical complexity of iterations.

\section{Non-zero correlation} \label{rhoNonZero}

 When $\rho(t) \neq 0$ we solve the PDE in \eqref{PDEx} by constructing a series in $\rho(t)$ around the point $\rho(t) = 0$. Thus, if the absolute value of the correlation is small for all $t \ge 0$, several terms in the series expansion could provide a suitable approximation to the exact solution. This kind of expansion has been already considered in the literature for some SV models with constant coefficients (see \citep{Antonelli2009} and references therein). It was shown there that this power series converges with positive radius under some regularity conditions.

Let us represent the solution of  \eqref{PDEx} as a series
\begin{equation} \label{series1}
u(t,x,\sigma) = \sum_{i=0}^\infty \rho^i(t) u_i(t,x,\sigma),
\end{equation}
\noindent where $\rho(t)$ is a small parameter. Using this representation in the zero-order approximation on $\rho(t)$ we get from \eqref{PDEx}
\begin{align} \label{pde0}
\fp{u_0}{t} &+ \sigma^2 \left[ \dfrac{1}{2} \sop{u_0}{x} + \frac{b}{x} \fp{u_0}{x} +  \dfrac{1}{2} \gamma^2(t) \sop{u_0}{\sigma} \right] - \kappa(t) \sigma \fp{u_0}{\sigma} = 0.
\end{align}
Let us also request that the first order approximation $u_0(t,x,\sigma)$ obeys the terminal condition in \eqref{tc1} and the boundary conditions  in \eqref{bc1}. Then, it is easy to see that such a problem coincides with that for the uncorrelated case. In Section~\ref{rho0sec} we obtained the solution of this problem which is given by \eqref{finSol1}.

In the first order approximation in $\rho(t)$ from \eqref{PDEx} we obtain the following PDE
\begin{align} \label{pde1}
\fp{u_1}{t} &+ \sigma^2 \left[ \dfrac{1}{2} \sop{u_1}{x} + \frac{b}{x} \fp{u_1}{x} +  \dfrac{1}{2} \gamma^2(t) \sop{u_1}{\sigma} \right] - \kappa(t) \sigma \fp{u_1}{\sigma} = - \gamma(t) \sigma^2 \cp{u_0}{x}{\sigma}.
\end{align}
As we have already satisfied the terminal and boundary conditions when constructing a zero-order approximation, this PDE should be solved subject to homogeneous terminal and boundary conditions.

 The \eqref{pde1} is an inhomogeneous PDE. Suppose that the Green's function $G_1(t,x|T,x_0) = G_1(x, x_0,T-t)$ of the homogeneous counterpart of \eqref{pde1} is known. Then,  by using the Duhamel's principle, \citep{Polyanin2002},  the solution of \eqref{pde1} reads
 \begin{equation} \label{callCIR1}
 u_1(t,x, \sigma) = \int_t^T dk \int_0^{y(k)} d\xi \int_{0}^\infty dz \, G_1(T-k,x,\sigma, \xi, z) \gamma(k) z^2 \cp{u_0(k,\xi,z)}{\xi}{z}.
 \end{equation}

 It is easy to see that next approximations can be constructed in a similar way to provide
 \begin{equation} \label{callCIR2}
 u_j(t,x, \sigma) = \int_t^T dk \int_0^{y(k)} d\xi \int_{0}^\infty dz \, G_j(T-k,x,\sigma, \xi, z) \gamma(k) z^2 \cp{u_{j-1}(k,\xi,z)}{\xi}{z}.
 \end{equation}

 Finally, observe that the homogeneous PDEs for all $u_j(t, x, \sigma), j \ge 0$ are same, i.e. $G_j = G_0, \ \forall j \ge 0$. That means that we need to find the Green's function only once. By definition, it solves the PDE in \eqref{pde0} with the boundary conditions in \eqref{bc1} and the terminal condition $G(T,x,\sigma, \xi, z) = \delta(x-\xi)$ where $\delta(x)$ is the Dirac Delta function. Thus, to find $G_0$ we can use the solution in \eqref{finSol1} by just replacing the terminal (initial in the variable $\tau$) condition.

\section{Numerical experiments} \label{numExp}

Here we aim to test the accuracy and speed of the proposed approach by comparing it with some FD method.
For running numerical tests an explicit form of the model parameters $\kappa(t), \gamma(t), \rho(t), \sigma, \beta$ should be specified. Since we wish to investigate the speed and accuracy of the method, there is no need to calibrate the model to real market quotes.

\begin{figure}[!htb]
\begin{center}
\hspace*{-0.3in}
\fbox{\includegraphics[width=0.6\textwidth]{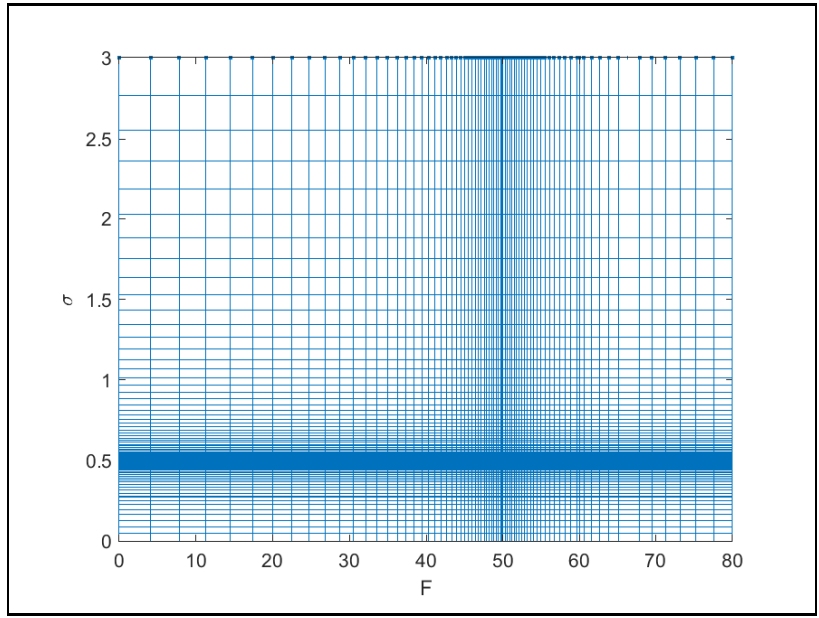}}
\end{center}
\caption{A typical nonuniform grid with 76 nodes in $F$ and 79 nodes in $\sigma$ for the FD scheme with $F_0 = 60, \ \sigma_0 = 0.5$ and $K=50$.}
\label{grid}
\end{figure}

We compare the barrier option prices obtained by using our method with those obtained by solving \eqref{PDE1} using the FD method described in detail in \citep{Itkin2014b}. This method belongs to a family of so-called ADI (alternative direction implicit) schemes and provides second order of approximation in all dimensions. Discretization of the PDE is done using a non-uniform grid which is compressed close to the initial forward price $F$ and the initial instantaneous volatility $\sigma$. A typical FD grid is presented in Fig.~\ref{grid}. The method starts with doing few Rannacher to provide better stability of the solution. The method has been validated for various problems. For instance, in \citep{CarrItkinMuravey2021} it was used to price an European vanilla Put in the Heston model with constant parameters $\kappa_0, \theta_0, \sigma_0, \rho_0, v_0$ since for this model the Put price can be found by FFT. It was observed that the relative error of the FD scheme in use is about 12 bps.

Also, for the Call option at $t=T$ there is a kink in the option price at the upper boundary $H$. Therefore, for a better stability it is useful to solve this problem for a covered Call ${\cal C}(t,F) = C(t,F) - F + K$. For the covered Call the boundary conditions obviously change to ${\cal C}(t,0) = K, \ {\cal C}(t,H) = K - H$ and the terminal condition - to
${\cal C}(T,F) = (F-K)^+ + K - F$. Thus, when the Call is In-The-Money at $t=T$, the payoff of the covered Call vanishes. Still at $F=H, \ t = T$ there is a kink in the payoff of the covered Call, but now of a much less amplitude. Also, the PDE for  ${\cal C}(t,F)$ coincides with that for $C(t,0)$ but now has an extra source term
$- r(F-K)$. This term, however, doesn't change the ADI scheme, and can be added to the explicit steps of the methods.

\paragraph{Example 1.} In this test we assume that the instantaneous volatility $\sigma$ is constant, i.e. not stochastic. Also let $y(t) = y - const, \ r(t) = r - const$. In this case \eqref{PDEv} can be solved analytically which yields
\begin{equation} \label{constSigAn}
\baru(t,\sigma,p) = \baru(T,p) e^{- \frac{1}{2} p^2 \sigma^2 (T-t)}.
\end{equation}
Thus, the Up-and-Out barrier Call option price can be expressed analytically as
\begin{equation} \label{CallTest1}
C(x,\sigma) = 2 e^{-r T} x^{-\nu} \sum_{n=1}^\infty \baru(T,\mu_n/y) e^{- \frac{\mu^2_n}{2 y^2} \sigma^2 T}
\frac{J_{|\nu|} \left( \frac{\mu_n x}{y}\right)}{J^2_{|\nu| + 1} (\mu_n)}.
\end{equation}

Note, that this expression can be also represented in a different form (despite, perhaps, less practical). Indeed, taking into account the definition of $\baru(T,p)$ as the GIT of the option payoff and substituting it into \eqref{CallTest1} we get
\begin{align} \label{CallPriceTheta}
C(x,\sigma) &= 2 e^{-r T} x^{1-2\nu} y \int_0^1 \left[A \xi^{-2 \nu}  - K\right]^+ \Theta_{|\nu|}(\sigma \sqrt{T}/y,\xi,\xi) d \xi, \qquad A = \left(- \beta  y\right)^{-1/\beta },
\end{align}
\noindent where the function $\Theta_{|\nu|}(\varsigma,x_1,x_2)$ was introduced in \citep{CarrItkinMuravey2020} as
\begin{equation}
\Theta_{|\nu|}(\varsigma,x_1,x_2) = \sum_{n=1}^\infty  e^{- \frac{\mu^2_n \varsigma}{2}}
\frac{x_1^\nu  J_{|\nu|} (\mu_n x_1)} {J_{|\nu| + 1} (\mu_n)} \frac{x_2^\nu J_{|\nu|} (\mu_n x_2)}{J_{|\nu| + 1} (\mu_n)}.
\end{equation}
This function (let us call it {\it the Bessel Theta function}) is a Bessel analog of the Jacobi Theta function of the $\theta_3(\phi,x)$, \citep{mumford1983tata}, in a sense, that $\theta_3(\phi,x)$ is a periodic solution of the heat equation, while $\Theta_{|\nu|}(\varsigma,x_1,x_2)$ is a periodic solution of the Bessel equation. It can be checked that, if $\nu = 1/2$, we have
\begin{align}
\Theta_{1/2}\left(\frac{\sqrt{\tau}}{y(\tau)}, \frac{s}{y(\tau)}, \frac{z}{y(\tau)} \right) =
\frac{y(\tau)}{ 2} \left[ \theta_3\left(e^{-\frac{\pi^2 \theta^2}{8}}, \frac{n \pi(s-z)}{4 y(\tau)}\right) -
\theta_3\left(e^{-\frac{\pi^2 \theta^2}{8}}, \frac{n \pi(s+z)}{4 y(\tau)}\right) \right].
\end{align}
Also, it can be checked that the Bessel Theta function at $\varsigma \to 0$ becomes the Dirac delta function. Therefore, $\Theta_{|\nu|}(\varsigma,x_1,x_2)$ in \eqref{CallPriceTheta} is a scaled density of the underlying process at the domain $F \in [0,y]$, and \eqref{CallPriceTheta} can be naturally treated as a discounted expectation of the option payoff under risk-neutral measure.

However, as we mentioned, representation \eqref{CallPriceTheta} is less practical. That is because the Jacobi theta functions are implemented in many numerical libraries and programming languages, like MpMath in python, or Wolfram Mathematica or Matlab, etc., while the Bessel Theta function is not. Therefore, manual implementation is required at the moment.

\begin{figure}[!htb]
\begin{center}
\hspace*{-0.3in}
\fbox{\includegraphics[width=0.6\textwidth]{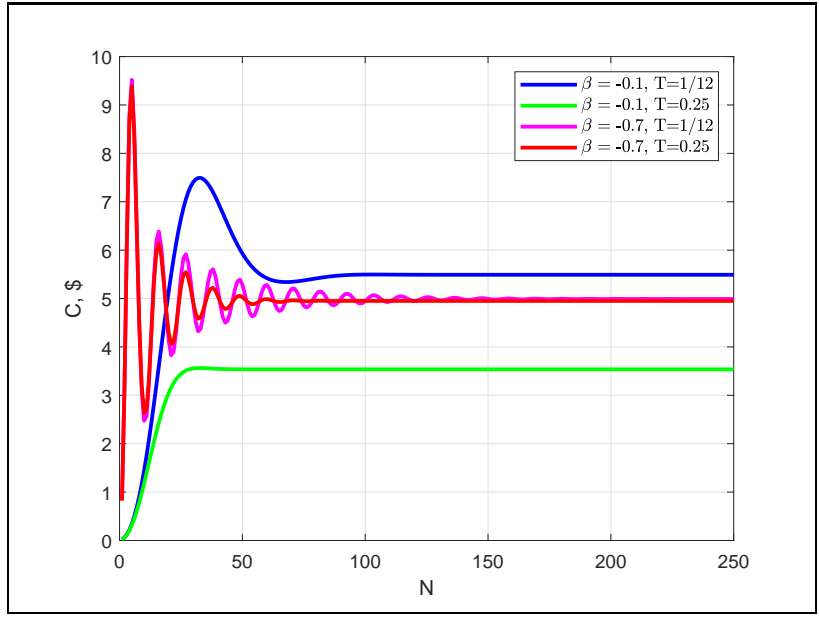}}
\end{center}
\caption{The Up-and-Out barrier Call option price computed by using \eqref{CallTest1} as a function of $N$ - the number of terms in the series.}
\label{convTest1}
\end{figure}

To proceed with the analytical representation of the Call option price in \eqref{CallTest1}, we take parameters of the model as in Table~\ref{paramTest1} and look at convergence of the option price as a function of the number of terms $N$ in \eqref{CallTest1}. The results are presented in Fig.~\ref{convTest1}. It can be seen that all series converge at $N \approx 150$ for short  maturities, and even faster for longer maturities.

\begin{table}[!htb]
\begin{center}
\begin{tabular}{|c|c|c|c|c|c|c|c|c|c|c|c|c|c|c|}
\hline
$F$ & $\sigma$ & $T$ & $r$ & $H$ & $K$ & $\beta$ \\
\hline
60 & 0.5 & 1/12, 0.25 & 0.02 & 80 & 55  & -0.1, -0.7\\
\hline
\end{tabular}
\caption{Parameters of the model in Example 1.}
\label{paramTest1}
\end{center}
\end{table}

In Table~\ref{fdTest1} we present the comparison of results computed by using \eqref{CallTest1} with $N=250$ with those obtained by using the FD method. As here the instantaneous volatility is constant and so the model becomes one-dimensional, we also provide the results obtained by using a 1D FD scheme described in \citep{CarrItkinMuravey2020} and also elapsed times of all methods.
\begin{table}[!htb]
  \centering
  \scalebox{0.9}{
  \begin{tabular}{|l|r|r|r|r|r|r|r|}
    \toprule
   \textbf{Method}  & \multicolumn{7}{c|}{\textbf{T, yrs}} \\
    \midrule
     & \textbf{0.042} & \textbf{0.083} & \textbf{0.250} & \textbf{0.500} & \textbf{1.000} & \textbf{2.000} & \multicolumn{1}{l|}{\textbf{El. time, sec}} \\
   \toprule
    \textbf{Analytic} & 5.1816 & 5.4908 & 5.0768 & 3.5365 & 1.8997 & 0.8333 & 0.007 \\
   \hline
    \textbf{FD 1D} & 5.1811 & 5.4875 & 5.0731 & 3.5352 & 1.8994 & 0.8332 & 0.17 \\
   \hline
    \textbf{FD 2D} & 5.1805 & 5.4813 & 5.0617 & 3.5299 & 1.8975 & 0.8326 & 1.60 \\
    \bottomrule
    \end{tabular}%
}
\caption{The Up-and-Out barrier Call option prices for $\sigma$ = const, $\beta$ = -0.1, computed by various methods and the elapsed time in secs (for one cell in the table).}
\label{fdTest1}%
\end{table}%
It can be seen that all prices obtained by various methods are close to each other while the analytic computation is obviously much faster.

Same results for $K=60$ are presented in Table~\ref{fdTest2}.
\begin{table}[!htb]
  \centering
  \scalebox{0.9}{
  \begin{tabular}{|l|r|r|r|r|r|r|r|}
    \toprule
   \textbf{Method}  & \multicolumn{7}{c|}{\textbf{T, yrs}} \\
    \midrule
     & \textbf{0.042} & \textbf{0.083} & \textbf{0.250} & \textbf{0.500} & \textbf{1.000} & \textbf{2.000} & \multicolumn{1}{l|}{\textbf{El. time, sec}} \\
   \toprule
    \textbf{Analytic} & 1.6203 & 2.2467 & 2.5756 & 1.8174 & 0.9565 & 0.4104 & 0.007 \\
   \hline
    \textbf{FD 1D} & 1.6172 & 2.2451 & 2.5729 & 1.8164 & 0.9562 & 0.4103 & 0.17 \\
   \hline
    \textbf{FD 2D} & 1.6182 & 2.2427 & 2.5647 & 1.8085 & 0.9476 & 0.4043 & 1.60 \\
    \bottomrule
    \end{tabular}%
}
\caption{The Up-and-Out barrier Call option prices for $\sigma$ = const, $\beta$ = -0.1 and $K=60$ (ATM) computed by various methods and the elapsed time in secs (for one cell in the table).}
\label{fdTest2}%
\end{table}%

\paragraph{Example 2.} Here we consider the whole $\lambda$-SABR model, so the instantaneous volatility is stochastic. Since we don't calibrate the model, without loss of generality we choose just some artificial time dependencies of the model parameters, namely:
\begin{equation} \label{ex}
\gamma(t) = \gamma_1 e^{-\gamma_2 t}, \quad \rho(t) = 0, \quad
\kappa(t) = \kappa_1 e^{-\kappa_2 t}, \quad r(t) = r_0, \quad H(t) = H.
\end{equation}
\noindent where $\gamma_1, \gamma_2, \kappa_1, \kappa_2$ are constants. With these definitions we can directly find
\begin{align}
\tau(t) &= \frac{\gamma_1^2}{4 \gamma_2} \left(e^{-2 \gamma_2 t}-e^{-2 \gamma_2 T}\right), \quad
g(t) = \frac{\kappa_1}{\kappa_2}\left(e^{-\kappa_2 T} - e^{-\kappa_2 t} \right) - \tau(t), \quad
\eta(t,p) = - \frac{p^2}{\gamma^2(t)} e^{-2 g(t)}.
\end{align}
Other parameters of the test are presented in Table~\ref{param}.
\begin{table}[!htb]
\begin{center}
\begin{tabular}{|c|c|c|c|c|c|c|c|c|c|c|c|c|c|c|}
\hline
$F$ & $\sigma$ & $\gamma_1$ & $\gamma_2$ & $\kappa_1$ & $\kappa_2$ & $H$   & $r_0$ \\
\hline
60 & 0.5 & 0.5 & 0.3 & 1. & 0.2 & 80 & 0.02  \\
\hline
\end{tabular}
\caption{Parameters of the model in Example 2.}
\label{param}
\end{center}
\end{table}
\vspace{-\baselineskip}

We run the test for a set of maturities $T \in [1/24, 1/12, 0.25,0.5,1,2]$ years and strikes $K \in [45, 50, 55, 60, 65, 70]$.  For the RBF method as the collocation points we use a uniform grid in $t \in [0,T]$ and $z \in [z_0 - z_m, z_0 + z_m]$ where $z_0 = \log \sigma + g(0)$, $z_m$ = 0.5. Close to ATM it is useful to add an extra (or even few) collocation point behind $\tau(0)$ for better approximation. We take $N_j = 10, N_l = 20$, $N = 350$. The best values of $\varepsilon$ in all experiments are given in Table~\ref{tabEps}. It can be seen that these values are almost independent of the strikes and maturities, as well as of $\beta$. Only for the ATM options we need to decrease $\varepsilon$\footnote{Since the payoff function has a kink ATM, it is always a problem for any numerical method to address it in the numerical solution. Various approached was developed in the literature, see a survey, e.g., in \citep{ItkinBook}.}.
%
\begin{table}[htbp]
  \centering
      \begin{tabular}{|r|r|r|r|r|r|r|}
    \toprule
    \multicolumn{1}{|c|}{$\mathbf \beta$} & \multicolumn{6}{c|}{K} \\
    \toprule
            & 45     & 50     & 55     & 60     & 65     & 70 \\
    \toprule
    -0.1  & 0.10  & 0.10  & 0.10  & 0.02  & 0.15  & 0.15 \\
    \hline
    -0.7  & 0.15  & 0.15  & 0.15  & 0.02  & 0.15  & 0.10 \\
    \bottomrule
    \end{tabular}%
\caption{The values of $\varepsilon$ used in the numerical experiments.}
  \label{tabEps}%
\end{table}%

When solving \eqref{VoltRBF} we approximate the integral in time by using a Simpson method with $N$ (in general, non-even) nodes distributed in $k \in [0,\tau]$.  For the FD method the time step is fixed and equal to $\min(0.01, T/50)$ year to preserve the method's accuracy in time at high $T$. Accordingly, since our FD scheme is of the second order of approximation in time, and the Simpson rule provides the fourth order of approximation, to preserve same accuracy using the Simpson method we need to take $N_j \approx 10 T$. This dictates our choice of $N_j = 10$ for all maturities (numerical experiments show that even for $T=2$ the increase of $N_j$ doesn't improve the results).

\begin{figure}[!htb]
\begin{center}
\hspace*{-0.3in}
\fbox{\includegraphics[width=0.6\textwidth]{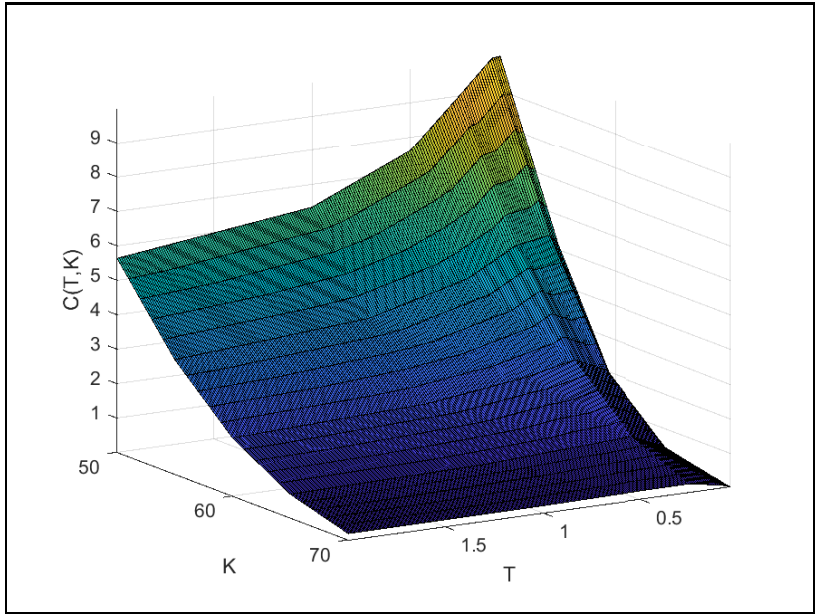}}
\end{center}
\caption{Up-and-Out Call option prices for the $\lambda$-SABR model obtained by the FD method in our experiment, with $\beta$ = -0.1.}
\label{callPriceFD}
\end{figure}

To solve \eqref{matSys} a standard \verb+bicgstabl+ iterative solver is used with no preconditioner. The Up-and-Out barrier Call option prices computed in these experiments using Matlab are presented in Tab.~\ref{TabComp} and also in Fig.~\ref{callPriceFD}. Typical elapsed times are also shown in Tab.~\ref{TabComp}. With $N=250$ the accuracy of the RBF method slightly drops down while the elapsed time becomes about 1.1 sec.
\begin{table}[!htb]
  \centering
  \scalebox{0.65}{
    \begin{tabular}{|c!{\vrule width 1.5pt} r|r|r|r|r|r!{\vrule width 1.5pt} r|r|r|r|r|r|}
    \specialrule{.1em}{.05em}{.05em}
  \fcolorbox{black}{green!30}{$\beta = -0.1$} \hfill $\bf{T}$     & 0.038 & 0.083 & 0.25  & 0.5   & 1     & 2     & 0.038 & 0.083 & 0.25  & 0.5   & 1     & 2 \\
    \specialrule{.1em}{.05em}{.05em}
    $\bf{K}$     & \multicolumn{6}{c!{\vrule width 1.5pt}}{\textbf{GIT}} & \multicolumn{6}{c|}{\textbf{FD}} \\
    \specialrule{.1em}{.05em}{.05em}
    70    & 0.0412 & 0.1151 & 0.2815 & 0.3452 & 0.2456 & 0.1997 & 0.0149 & 0.0992 & 0.2713 & 0.2635 & 0.2255 & 0.2005 \\ \hline
    65    & 0.3286 & 0.6547 & 0.9666 & 0.9584 & 0.6648 & 0.6847 & 0.2280 & 0.5799 & 1.0118 & 0.9352 & 0.7918 & 0.7027 \\ \hline
    60    & 1.7614	& 2.2653	& 2.6923	& 2.3719	& 1.8738	& 1.7136 & 1.5838 & 2.1574 & 2.5960 & 2.2809 & 1.9143 & 1.6972 \\ \hline
    55    & 5.1455 & 5.3522 & 5.2544 & 4.4732 & 3.6379 & 3.3626 & 5.1691 & 5.4385 & 5.2927 & 4.5130 & 3.7610 & 3.3325 \\ \hline
    50    & 9.7614 & 9.6848 & 9.1691 & 7.4481 & 6.1788 & 5.6118 & 9.9943 & 9.9678 & 9.0054 & 7.5949 & 6.3409 & 5.6325 \\ \hline
    45    & 14.6288 & 14.4850 & 13.2513 & 11.0707 & 9.2756 & 8.4581 & 14.9845 & 14.8887 & 13.3691 & 11.3469 & 9.5560 & 8.5304 \\
\specialrule{.1em}{.05em}{.05em}
Elapsed time & 1.5      & 1.5      & 1.5      & 1.5      & 1.5      & 1.5  & 1.5  & 1.5  & 1.5  & 1.5   & 3.1   & 6.1 \\
    \specialrule{.1em}{.05em}{.05em}
 \fcolorbox{black}{green!30}{$\beta = -0.7$} \hfill $\bf{T}$     & 0.038 & 0.083 & 0.25  & 0.5   & 1     & 2     & 0.038 & 0.083 & 0.25  & 0.5   & 1     & 2 \\ \hline
    \specialrule{.1em}{.05em}{.05em}
    $\bf{K}$     & \multicolumn{6}{c!{\vrule width 1.5pt}}{\textbf{GIT}} & \multicolumn{6}{c|}{\textbf{FD}} \\
    \specialrule{.1em}{.05em}{.05em}
        70    & 0.0038 & 0.0136 & 0.0131 & 0.0254 & 0.0126 & 0.0027 & 6.9416E-08 & 2.7755E-07 & 2.4938E-06 & 1.0001E-05 & 2.1420E-05 & 5.1659E-05 \\ \hline
    65    & 0.0130 & 0.0156 & 0.0237 & 0.0146 & 0.0271 & 0.0037 & 3.4708E-08 & 1.3877E-07 & 1.3181E-06 & 1.6641E-05 & 2.0137E-04 & 6.7151E-04 \\ \hline
    60    & 0.2432	& 0.2854	& 0.3486	& 0.4184	& 0.4538	& 0.4711 & 0.1258 & 0.1806 & 0.2968 & 0.3800 & 0.4494 & 0.4820 \\ \hline
    55    & 5.0148 & 4.9903 & 4.9794 & 4.9982 & 4.9096 & 4.7928 & 4.9958 & 4.9917 & 4.9751 & 4.9503 & 4.9011 & 4.8045 \\ \hline
    50    & 9.9764 & 9.9895 & 9.9332 & 9.8322 & 9.7938 & 9.6076 & 9.9917 & 9.9833 & 9.9501 & 9.9005 & 9.8020 & 9.6079 \\ \hline
    45    & 14.9749 & 14.9291 & 14.8793 & 14.8485 & 14.7052 & 14.4162 & 14.9875 & 14.9750 & 14.9252 & 14.8507 & 14.7030 & 14.4118 \\ \hline
    \specialrule{.1em}{.05em}{.05em}
    Elapsed time & 1.5      & 1.5      & 1.5      & 1.5      & 1.5      & 1.5   & 1.6  & 1.4  & 1.4  & 1.4   & 2.7   & 5.7 \\
    \specialrule{.1em}{.05em}{.05em}
    \end{tabular}%
  }
  \caption{Comparison of Up-and-Out Call option prices for the $\lambda$-SABR model obtained by various methods and elapsed time in secs.}
  \label{TabComp}%
\end{table}%

\begin{figure}[!htb]
\begin{center}
\hspace*{-0.3in}
\fbox{\includegraphics[width=0.6\textwidth]{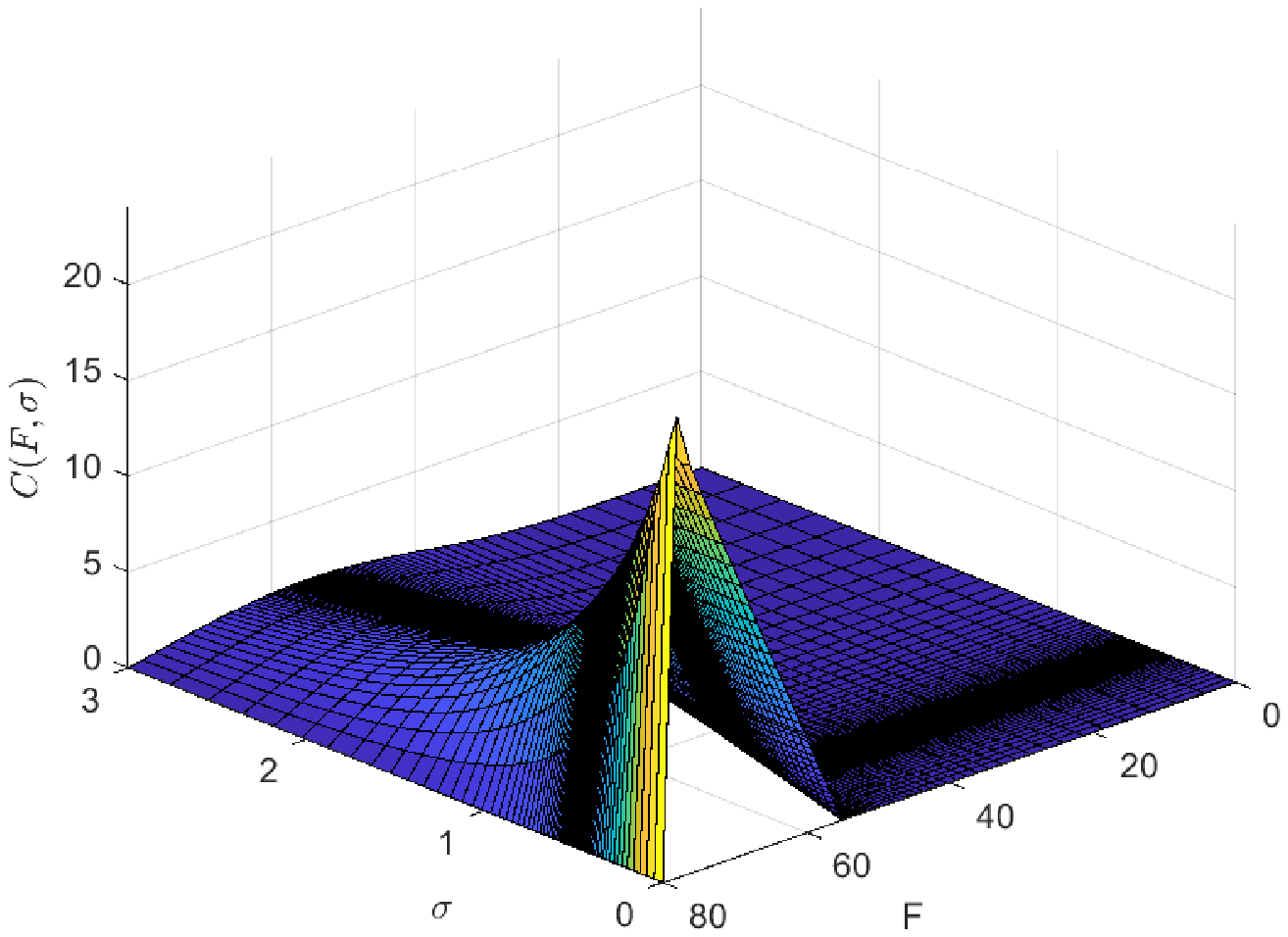}}
\end{center}
\caption{Up-and-Out barrier Call option $C(F,\sigma)$ for $\beta=-0.1, \ K = 55, \ T=1/24$, and other model parameters as in Table.~\ref{TabComp}.}
\label{surface}
\end{figure}

Fig.~\ref{surface} presents a typical Up-and-Out barrier Call option surface $C(F,\sigma)$ obtained for $\beta=-0.1, \ K = 55, \ T=1/24$, and other model parameters as in Table.~\ref{TabComp}.

In Table~\ref{errorRBF-FD} the percentage error between the option prices obtained by our and the FD method are presented as a function of $K, T$.
\begin{table}[htbp]
  \centering
  \small
    \begin{tabular}{|c|r|r|r|r|r|r|}
    \toprule
        $\beta = -0.1$  & \multicolumn{6}{c|}{\textbf{T}} \\
    \toprule
    \textbf{K} & 0.0380 & 0.0830 & 0.2500 & 0.5000 & 1.0000 & 2.0000 \\
    \toprule
    70    & 63.74 & 13.78 & 3.61  & 23.66 & 8.20  & -0.41 \\
    \hline
    65    & 30.63 & 11.42 & -4.67 & 2.42  & -19.09 & -2.63 \\
    \hline
    60    & 10.08	& 4.76	& 3.58	& 3.84	& -2.16	& 0.96  \\
    \hline
    55    & -0.46 & -1.61 & -0.73 & -0.89 & -3.38 & 0.89 \\
    \hline
    50    & -2.39 & -2.92 & 1.78  & -1.97 & -2.62 & -0.37 \\
    \hline
    45    & -2.43 & -2.79 & -0.89 & -2.50 & -3.02 & -0.85 \\
    \bottomrule
        $\beta = -0.7$  & \multicolumn{6}{c|}{\textbf{T}} \\
    \toprule
    \textbf{K} & 0.0380 & 0.0830 & 0.2500 & 0.5000 & 1.0000 & 2.0000 \\
    \toprule
    70    & 100.00 & 100.00 & 99.98 & 99.96 & 99.83 & 98.10 \\
    \hline
    65    & 100.00 & 100.00 & 99.99 & 99.89 & 99.26 & 81.93 \\
    \hline
    60    & 48.29	& 36.74 & 	14.88	& 9.16	& 0.98	& -2.30 \\
    \hline
    55    & 0.38  & -0.03 & 0.09  & 0.96  & 0.17  & -0.24 \\
    \hline
    50    & -0.15 & 0.06  & -0.17 & -0.69 & -0.08 & 0.00 \\
    \hline
    45    & -0.08 & -0.31 & -0.31 & -0.01 & 0.02  & 0.03 \\
    \bottomrule
    \end{tabular}%
  \caption{The percentage error between the option prices obtained by our and the FD method as a function of $K, T$. }
  \label{errorRBF-FD}%
\end{table}%
Big relative error at $K=70,65$ doesn't seem to confuse readers since the option price itself is very small (several cents at $K=70$). However, the ATM prices have bigger error. We attribute this error to both the simplistic numerical approach we use to solve the LMVF equation, and to the accuracy of the FD solver for the ATM prices. This can be seen from Tables~\ref{fdTest1},\ref{fdTest2} where the difference between the ATM prices computed by our method with the exact solution for $\baru$ (when $\gamma(t) = \kappa(t) = 0$) and those obtained by using the FD methods also slightly increases as compared with the OTM prices.

The agreement between various methods is much better for the ITM prices. This is because those are closer to the intrinsic value and so the values of the integrals are small (and so are the errors). Also, the errors are bigger at small maturities and lower at the high ones. A partial explanation of this effect was provided in \citep{CarrItkinMuravey2020}, where the authors compared the GIT and HP (heat potential) methods as applied to pricing barrier options under one-dimensional CIR and CEV models. It was observed that the GIT method provides very accurate results at high maturities in contrast to the HP method which is good at small $T$. This can be verified by looking at exponents under the GIT solution integral (see, e.g., \eqref{constSigAn}) which are {\it proportional} to $T$. Contrary, for short maturities the GIT method is slightly less accurate than the HP method, as the exponents in the HP  solution integral are {\it inversely proportional} to $T$.

Another possible reason could be different properties of the RBF matrix at various maturities. To check that we recomputed the ATM prices by using a \verb+minres+ solver which is good when the matrix is not positive definite, but symmetric, and we can still construct an orthogonal basis for the Krylov subspace by three-term recurrence relations, \citep{minres1975}. Indeed, we do see that due to small rounding errors some eigenvalues of our RBF matrices are very small negative numbers, therefore, in principle, the RBF matrix should be first regularized by using a standard procedure.  However, again, since this error depends on quality of the RBF interpolation it is expected that modern methods which are more stable than the Gaussian global RBF method would provide better accuracy. Indeed, for the Gaussian RBF method a typical matrix in \eqref{matSys}  has the condition number about $10^{19}$ and, therefore, even iterative methods of solving this system of linear equations produce bigger error than in the case of a well-conditioned matrix.

We also double the value of $N$ when using \verb+minres+ at $\beta=-0.7$ while doing so for \verb+bicgstabl+ doesn't change the computed prices. The results of these tests for the ATM options are presented in Table~\ref{ATMres}.  It can be seen that using the \verb+minres+ solver significantly drops down the relative error as compared with the \verb+bicgstabl+ solver at small and intermediate maturities, while the later solver outperforms  the former at high maturities. Also \verb+minres+ is almost twice faster than \verb+bicgstabl+, and so twice faster than the FD solver at small and intermediate $T$.
\begin{table}[!htb]
  \centering
  \scalebox{0.78}{
  \begin{tabular}{|l|rrrrrr|rrrrrr|}
    \toprule
    $\beta$ / solver & \multicolumn{6}{c|}{\textbf{T, C}}            & \multicolumn{6}{c|}{\textbf{T, \% dif}} \\
\cmidrule{2-13}          & 0.0380 & 0.0830 & 0.2500 & 0.5000 & 1.0000 & 2.0000 & 0.0380 & 0.0830 & 0.2500 & 0.5000 & 1.0000 & 2.0000 \\
    \toprule
    -0.7/bicgstabl & 0.2432 & 0.2596 & 0.3634 & 0.4055 & 0.4094 & 0.3919 & 48.2942 & 30.4523 & 18.3371 & 6.2793 & -9.7604 & -22.9764 \\
    -0.7/minres & 0.1503 & 0.1869 & 0.2644 & 0.3429 & 0.3685 & 0.3331 & 16.3537 & 3.3986 & -12.2616 & -10.8419 & -21.9523 & -44.6957 \\
    \toprule
    -0.1/bicgstabl & 1.7614 & 2.2653 & 2.6923 & 2.3719 & 1.8738 & 1.7136 & 10.0814 & 4.7629 & 3.5769 & 3.8370 & -2.1570 & 0.9576 \\
    -0.1/minres & 1.6073 & 2.0037 & 2.2804 & 2.2098 & 2.1352 & 2.1290 & 1.4603 & -7.6687 & -13.8411 & -3.2192 & 10.3477 & 20.2828 \\
    \bottomrule
    \end{tabular}%
    }
\caption{The Up-and-Out ATM Call option price computed by using two different solvers for $\beta$ = -0.7, -0.1 (left panel) and the percentage difference between these prices and those computed by using the FD method (right panel).}
  \label{ATMres}%
\end{table}%

\section{Discussion} \label{Discus}

In this paper we extend the GIT method proposed to get semi-analytical prices of barrier options for various 1D time-dependent models with time-dependent barriers, see \citep{ItkinLiptonMuraveyBook} and references therein,
by applying it to the so-called $\lambda$-SABR stochastic volatility model. In doing so we develop a new modification of the GIT method  which delivers solution of this problem as a Fourier--Bessel series. Coefficients of this series solve a LMVF equation of the second kind which is derived and discussed in the paper.

Let us compare this approach step-by-step with the numerical FD method.
\begin{enumerate}
\item With the FD method we create a 3D grid in $x-\sigma-t$ space, so the complexity of the solution is about $O(N_t N_x N_\sigma)$, where $N_i$ is the corresponding number of nodes in the $i$-th direction. With the GIT method we still some nodes in the $t$ and $\sigma$ space. Since integration in the $t$ space can be done by using high order quadratures, the number of the RBF nodes $n_t$ can be significantly reduced providing same accuracy. Thus, in general $n_t \ll N_t$. In the $\sigma$ space the GIT method doesn't require numerical integration (it is done analytically), and thus, we typically have $n_\sigma \ll N_\sigma$.

\item The GIT method doesn't need any grid in $x$. Instead, it needs to compute multiple terms in the Fourier-Bessel series. Fortunately, all matrices in \eqref{matSys} can be precomputed and reused. Therefore, solution of these systems for various $\mu_i, i=1,\ldots,$ is fast. Since the series converges, it can be truncated to some fixed number of terms.

    Also, all terms in this series are produced by solving the corresponding LMVF equation which are independent of each other. Therefore, they can be solved in parallel by using any parallel architecture.

\item We compared pricing of Call barrier options by using the GIT and FD methods strike-by-strike and maturity-by-maturity. In other words, we used a backward approach. If one needs prices for multiple $K$ and $T$ a better way would be to solve the forward equation for the density of the underlying process and then integrate the solution with payoffs thus computing the discounted expectation. It turns out that the proposed GIT method is a perfect fit for doing so. Indeed, when solving the LMVF equation we can take $\tau$ to be $\tau(\max(T))$. Since our method uses the RBF interpolation, it generates the results for all $\tau \in [0,\tau(\max(T))]$ just in one sweep. Therefore, multiple $T$ can be priced at once. For multiple strikes, assuming that $\varepsilon$ is independent on $K$, the only remaining function of $K$ is $\baru(T,p)$ which enters the RHS of \eqref{matSys}. Therefore, it can be very fast precomputed for multiple strikes (just a vector operation), and then we can solve \eqref{matSys} for various RHS simultaneously. Indeed, it requires just solving one system with multiple RHS sides that could be efficiently done with the modern software. Thus, prices for all pairs $T,K$ can be obtained in one sweep. Therefore, the relative performance of the forward FD and GIT methods remains the same (the FD method also requires computing multiple expectations, i.e. the integrals with the lower limit being equal to $K$).

\item Based on the results obtained in the paper, performance of the GIT method is almost twice better than that of the FD method at small maturities (by using the \verb+minres+ solver), of the same order at intermediate  maturities (by using the \verb+bicgstabl+ solver), while the GIT method is faster at long maturities. The accuracy of the GIT method combined together with the Gaussian RBF is good but depends on the value of the shape parameter $\varepsilon$ (as it should be) and a choice of the linear solver. We show that choosing an appropriate solver gives rise to good agreement between our and FD benchmark results.

\item To solve the LMVF equations we used the Gaussian RBF method. As mentioned in Section~\ref{sec:RBF}, modern RBF approaches significantly improve both speed and accuracy of the solution, and, therefore, it is expected that using these methods will significantly improve both speed and accuracy. This is subject of our future investigation which we will report elsewhere.

\end{enumerate}

On a general note, solving a problem with time-dependent barriers by using a FD method brings some technical challenges. In this case either the FD gird has to be reconstructed at every time step to have the barrier as the last node, or a fixed grid can be used but the boundary condition should be moved along the grid, e.g., by interpolating the barrier onto the closest node. The later decreases the accuracy of the method while the former deteriorates the speed. In contrast, our semi-analytical method doesn't face this problem at all.

Another advantage of our method as compared with the FD method is that it is semi-analytical. This means, that various properties of the solution can be retrieved by analyzing its Fourier-Bessel series representation. For instance, computation of Greeks can be done by differentiating this series by the necessary parameter of the model. In particular, derivatives on $x$ (Delta, Gamma, etc.) come immediately for free as well as various derivatives on $\sigma$.

A weak point of the proposed method for the SABR model (but not for the Heston model as this is shown in \citep{CarrItkinMuravey2021}) is taking into account correlation between two Brownian motions of the model.  We developed our approach assuming zero correlation, and then showed how small correlations could be taken into account. However, for strong correlations the asymptotic approach is not feasible. Therefore, at the moment this remains to be an open question how the GIT method could be applied to the $\lambda$-SABR model with strong correlation. One idea is to switch dimensions, i.e. to construct the GIT in the $\sigma$ space first and then it can be explicitly inverted as the model is log-normal in $\sigma$, \citep{ItkinLiptonMuraveyBook}. Then instead of the LMVF equation in $t-\sigma$ variables we obtain a similar LMVF equation in $t-x$ variables which now includes correlation. Solving it we can obtain the option price which in this case is represented not via series, but is just proportional to the solution of the LMVF equation. We will investigate this approach in our future work.

\section*{Acknowledgments}

We are grateful to Peter Carr, Elisabeth Larsson, Alexander Lipton and Fazlollah Soleymani for various discussions. Dmitry Muravey acknowledges support by the Russian Science Foundation under the Grant number 20-68-47030.



\begin{thebibliography}{40}
\providecommand{\natexlab}[1]{#1}
\providecommand{\url}[1]{\texttt{#1}}
\expandafter\ifx\csname urlstyle\endcsname\relax
  \providecommand{\doi}[1]{doi: #1}\else
  \providecommand{\doi}{doi: \begingroup \urlstyle{rm}\Url}\fi

\bibitem[Abramowitz and Stegun(1964)]{as64}
M.~Abramowitz and I.~Stegun.
\newblock \emph{Handbook of Mathematical Functions}.
\newblock Dover Publications, Inc., 1964.

\bibitem[Antonelli and Scarlatti(2009)]{Antonelli2009}
F.~Antonelli and S.~Scarlatti.
\newblock Pricing options under stochastic volatility:a power series approach.
\newblock \emph{Finance Stoch}, 13:\penalty0 269--303, 2009.

\bibitem[Antonov et~al.(2019)Antonov, Konikov, and Spector]{Antonov2019}
A.~Antonov, M.~Konikov, and M.~Spector.
\newblock \emph{Modern SABR Analytics}.
\newblock SpringerBriefs in Quantitative Finance. Springer, 2019.
\newblock ISBN 978-3-030-10656-0.

\bibitem[Assari et~al.(2019)Assari, Asadi-Mehregan, and Dehghan]{Assari2019}
P.~Assari, F.~Asadi-Mehregan, and M.~Dehghan.
\newblock {On the numerical solution of Fredholm integral equations utilizing
  the local radial basis function method}.
\newblock \emph{International Journal of Computer Mathematics}, 96\penalty0
  (7):\penalty0 1416--1443, 2019.

\bibitem[Barger and Lorig(2017)]{BargerLorig2017}
W.~Barger and M.~Lorig.
\newblock Approximate pricing of european and barrier claims in a
  local-stochastic volatility setting.
\newblock \emph{International Journal of Financial Engineering}, 4\penalty0
  (02n03):\penalty0 1750017, 2017.

\bibitem[Bateman and Erd{\'e}lyi(1953)]{bateman1953higher}
H.~Bateman and A.~Erd{\'e}lyi.
\newblock \emph{Higher Transcendental Functions}, volume~1 of \emph{Bateman
  Manuscript Project California Institute of Technology}.
\newblock McGraw-Hill, 1953.

\bibitem[Carr and Itkin(2021)]{CarrItkin2020jd}
P.~Carr and A.~Itkin.
\newblock {Semi-closed form solutions for barrier and American options written
  on a time-dependent Ornstein-Uhlenbeck process}.
\newblock \emph{Journal of Derivatives}, 29\penalty0 (1):\penalty0 9--26, 2021.

\bibitem[Carr and Sun(2007)]{CarrSun}
P.~Carr and J.~Sun.
\newblock A new approach for option pricing under stochastic volatility.
\newblock \emph{Review of Derivatives Research}, 10:\penalty0 87--250, 2007.

\bibitem[Carr et~al.(2020)Carr, Itkin, and Muravey]{CarrItkinMuravey2020}
P.~Carr, A.~Itkin, and D.~Muravey.
\newblock {Semi-closed form prices of barrier options in the time-dependent CEV
  and CIR models}.
\newblock \emph{Journal of Derivatives}, 28\penalty0 (1):\penalty0 26--50,
  2020.

\bibitem[Carr et~al.(2021)Carr, Itkin, and Muravey]{CarrItkinMuravey2021}
P.~Carr, A.~Itkin, and D.~Muravey.
\newblock Semi-analytical pricing of barrier options in the time-dependent
  {Heston} model.
\newblock In preparation, 2021.

\bibitem[Fasshauer and McCourt(2012)]{McCourt2012}
G.E. Fasshauer and M.J. McCourt.
\newblock Stable evaluation of {Gaussian Radial Basis} function interpolants.
\newblock \emph{SIAM Journal on Scientific Computing}, 34:\penalty0 A737--A762,
  2012.

\bibitem[Fornberg et~al.(2011)Fornberg, Larsson, and Flyer]{Fornberg2}
B.~Fornberg, E.~Larsson, and N.~Flyer.
\newblock Stable computations with {Gaussian Radial Basis} functions.
\newblock \emph{SIAM J. Sci. Comput.}, 33\penalty0 (2):\penalty0 869--892,
  2011.

\bibitem[Gorovoi and Linetsky(2004)]{GorovoiLinetsky}
V.~Gorovoi and V~Linetsky.
\newblock Black's model of interest rates as options, eigenfunction expansions
  and japanese interest rates.
\newblock \emph{Mathematical Finance}, 14\penalty0 (1):\penalty0 49--78, 2004.

\bibitem[Gray and Pinsky(1992)]{Pinsky1992}
A.~Gray and M.A. Pinsky.
\newblock {Computer graphics and a new Gibbs phenomenon for Fourier-Bessel
  series}.
\newblock \emph{Experimental Mathematics}, 1\penalty0 (4):\penalty0 313--316,
  1992.

\bibitem[Hagan et~al.(2002)Hagan, Kumar, A, and Woodward]{hagan2002}
P.~Hagan, D.~Kumar, A.~Lesniewski A, and D~Woodward.
\newblock Managing smile risk.
\newblock \emph{Wilmott magazine}, pages 84--108, September 2002.

\bibitem[Hagan et~al.(2014)Hagan, Kumar, Lesniewski, and Woodward]{SABR2014}
P.S. Hagan, D.~Kumar, A.~Lesniewski, and D.~Woodward.
\newblock Arbitrage-free {SABR}.
\newblock \emph{Wilmott}, pages 60--75, January 2014.

\bibitem[Hagan et~al.(2020)Hagan, Lesniewski, and Woodward]{Hagan2020}
P.S Hagan, A.~Lesniewski, and D.E Woodward.
\newblock Implied volatilities for mean reverting {SABR} models.
\newblock \emph{Wilmott}, pages 62--77, July 2020.

\bibitem[Hardy(1991)]{Hardy1991}
G.H. Hardy.
\newblock \emph{Divergent Series}.
\newblock AMS Chelsea Publishing, 2 edition, 1991.
\newblock ISBN 978-0821826492.

\bibitem[Henry-Labordere(2005)]{Labordere2005}
P.~Henry-Labordere.
\newblock A general asymptotic implied volatility for stochastic volatility
  models, 2005.
\newblock URL \url{https://arxiv.org/pdf/cond-mat/0504317.pdf}.

\bibitem[Itkin(2015)]{Itkin2014b}
A.~Itkin.
\newblock {High-Order Splitting Methods for Forward PDEs and PIDEs}.
\newblock \emph{International Journal of Theoretical and Applied Finance},
  18\penalty0 (5):\penalty0 1550031--1 ---1550031--24, 2015.

\bibitem[Itkin(2017)]{ItkinBook}
A.~Itkin.
\newblock \emph{Pricing derivatives under {L{\'e}vy} models}.
\newblock Number~12 in Pseudo-Differential Operators. Birkhauser, Basel, 1
  edition, 2017.

\bibitem[Itkin and Muravey(2021)]{ItkinMuraveyDBFMF}
A.~Itkin and D.~Muravey.
\newblock Semi-analytic pricing of double barrier options with time-dependent
  barriers and rebates at hit.
\newblock \emph{Frontiers of Mathematical Finance}, 1:\penalty0 1--36, 2021.

\bibitem[Itkin et~al.(2020)Itkin, Lipton, and Muravey]{ItkinLiptonMuravey2020}
A~Itkin, A.~Lipton, and D.~Muravey.
\newblock {From the Black-Karasinski to the Verhulst model to accommodate the
  unconventional Fed's policy}, June 2020.
\newblock URL \url{https://arxiv.org/abs/2006.11976}.

\bibitem[Itkin et~al.(2021)Itkin, Lipton, and Muravey]{ItkinLiptonMuraveyBook}
A.~Itkin, A.~Lipton, and D.~Muravey.
\newblock \emph{Generalized Integral Transforms in Mathematical Finance}.
\newblock WSPC, Singapore, 2021.
\newblock ISBN 978-981-123-173-5.

\bibitem[Kartashov(2001)]{kartashov2001}
E.M. Kartashov.
\newblock \emph{Analytical Methods in the Theory of Heat Conduction in Solids}.
\newblock Vysshaya Shkola, Moscow, 2001.

\bibitem[Kato et~al.(2013)Kato, Takahashi, and Yamada]{Kato2013}
T.~Kato, A.~Takahashi, and T.~Yamada.
\newblock {An asymptotic expansion formula for Up-and-Out barrier option price
  under stochastic volatility model}.
\newblock \emph{JSIAM Letters}, 5:\penalty0 17--20, 2013.

\bibitem[Larsson and Fornberg(2005)]{Larsson7}
E.~Larsson and B.~Fornberg.
\newblock Theoretical and computational aspects of multivariate interpolation
  with increasingly flat radial basis functions.
\newblock \emph{Comput. Math. Appl.}, 49\penalty0 (1):\penalty0 103--130, 2005.

\bibitem[Larsson et~al.(2013)Larsson, Lehto, Heryudono, and Fornberg]{Larsson2}
E.~Larsson, E.~Lehto, A.~Heryudono, and B.~Fornberg.
\newblock Stable computation of differentiation matrices and scattered node
  stencils based on {G}aussian radial basis functions.
\newblock \emph{SIAM J. Sci. Comput.}, 35\penalty0 (4):\penalty0 2096--2119,
  2013.

\bibitem[Lipton(2001)]{Lipton2001}
A.~Lipton.
\newblock \emph{Mathematical Methods For Foreign Exchange: A Financial
  Engineer's Approach}.
\newblock World Scientific, 2001.

\bibitem[Mumford et~al.(1983)Mumford, Nori, Previato, and
  Stillman]{mumford1983tata}
D.~Mumford, C.~Musiliand~M. Nori, E.~Previato, and M.~Stillman.
\newblock \emph{Tata Lectures on Theta}.
\newblock Progress in Mathematics. Birkh{\"a}user Boston, 1983.
\newblock ISBN 9780817631093.

\bibitem[Paige and Saunders(1975)]{minres1975}
C.~Paige and M.~Saunders.
\newblock Solution of sparse indefinite systems of linear equations.
\newblock \emph{{SIAM J. Numer. Anal.}}, 12:\penalty0 617--629, 1975.

\bibitem[Platen(1997)]{Platen1997}
E.~Platen.
\newblock A non-linear stochastic volatility model.
\newblock Financial Mathematics Research Report FMRR005-97, Canberra: Center
  for Financial Mathematics, Australian National University, 1997.

\bibitem[Polyanin(2002)]{Polyanin2002}
A.D. Polyanin.
\newblock \emph{Handbook of linear partial differential equations for engineers
  and scientists}.
\newblock Chapman \& Hall/CRC, 2002.

\bibitem[Revuz and Yor(1999)]{RevuzYor1999}
D.~Revuz and M.~Yor.
\newblock \emph{Continuous Martingales and Brownian Motion}.
\newblock Springer, Berlin, Germany, 3rd edition, 1999.

\bibitem[Shreve(1992)]{Shreve:1992}
S.~Shreve.
\newblock Martingales and the theory of capital-asset pricing.
\newblock \emph{Lecture Notes in Control and Information SCIENCES},
  180:\penalty0 809--823, 1992.

\bibitem[Thakoor et~al.(2019)Thakoor, Tangman, and Bhuruth]{Thakoor2019}
N.~Thakoor, D.Y. Tangman, and M.A. Bhuruth.
\newblock A spectral approach to pricing of arbitrage-free {SABR} discrete
  barrier options.
\newblock \emph{Computational Economics}, 54:\penalty0 1085--1111, 2019.

\bibitem[{Van der Stoep} et~al.(2015){Van der Stoep}, Grzelak, and
  Oosterlee]{Oosterlee2015}
A.W. {Van der Stoep}, L.A. Grzelak, and C.W. Oosterlee.
\newblock The time-dependent {FX-SABR} model: Efficient calibration based on
  effective parameters.
\newblock \emph{International Journal of Theoretical and Applied Finance},
  18\penalty0 (6):\penalty0 1550042, 2015.

\bibitem[Watson(1966)]{Watson1966}
G.N. Watson.
\newblock \emph{A Treatise on the Theory of Bessel Functions}.
\newblock Cambridge University Press, Cambridge, UK, 2nd edition, 1966.

\bibitem[Yang et~al.(2017)Yang, Liu, and Cui]{YangLiuCui2017}
N.~Yang, Y.~Liu, and Z.~Cui.
\newblock Pricing continuously monitored barrier options under the {SABR}
  model: A closed form approximation.
\newblock \emph{JMSE}, 2\penalty0 (2):\penalty0 116--131, 2017.

\bibitem[Zhang et~al.(2014)Zhang, Chen, and Nie]{Zhang2014}
H.~Zhang, Y.~Chen, and X.~Nie.
\newblock Solving the linear integral equations based on radial basis function
  interpolation.
\newblock \emph{Journal of Applied Mathematics}, 2014:\penalty0 793582, 2014.

\end{thebibliography}

\end{document}